\documentclass[preprint,5p,times]{elsarticle}
\usepackage[T1]{fontenc}
\usepackage{float}
\usepackage{amsmath}
\usepackage{amsthm}
\usepackage{bm}
\usepackage{breakurl}
\usepackage{breqn}
\usepackage[breaklinks]{hyperref}
\usepackage{amssymb}
\usepackage{array} % for defining a new column type
\usepackage{graphicx} % for \includegraphics
\usepackage{varwidth} %for the varwidth minipage environment
\usepackage{listings}
\usepackage{gensymb}
\usepackage{fancyhdr}
\usepackage{url}
\usepackage{lineno}
\usepackage[table]{xcolor}
\usepackage{colortbl}
\usepackage{multirow}
\usepackage{algorithm}% http://ctan.org/pkg/algorithms
\usepackage{algpseudocode}% http://ctan.org/pkg/algorithmicx
\usepackage{mathtools}
\usepackage{esdiff}

\usepackage{array}          % for !{\vrule width ...}

\definecolor{FastTeal}{RGB}{0,150,136}
\newcommand{\thickhline}{\noalign{\hrule height 1.8pt}}

\algnewcommand{\LeftComment}[1]{\Statex \(\triangleright\) #1}

\definecolor{temporal}{RGB}{112, 186, 164}
\definecolor{red}{RGB}{219, 191, 186}
\definecolor{orange}{RGB}{191, 167, 214}
\definecolor{yellow}{RGB}{240, 230, 140}

\journal{Future Generation Computer Systems}

\begin{document}
\begin{frontmatter}

\title{Advancing RT Core-Accelerated Fixed-Radius Nearest Neighbor Search}

\author[a]{Enzo Meneses}
\author[b]{Hugo Bec}
\author[a]{Crist\'obal A. Navarro\corref{author}}
\author[b]{Beno{\^i}t Crespin}
\author[a]{Felipe A. Quezada}
\author[c]{Nancy Hitschfeld}
\author[b]{Heinich Porro}
\author[b]{Maxime Maria}
\cortext[author] {Corresponding author.\\\textit{E-mail address:} cristobal.navarro@uach.cl}
\address[a]{Instituto de Informática, Universidad Austral de Chile.}
\address[b]{Univ. of Limoges, XLIM UMR 7252, CNRS, France}
\address[c]{Computer Science Department (DCC), University of Chile.}

\begin{abstract}
Recent research on GPU Computing has found new ways to leverage the Ray Tracing (RT) cores beyond lighting/rendering tasks. One relevant case of success is the acceleration of the fixed-radius nearest neighbors (FRNN) search for physical simulations. Although this idea of finding neighbors with GPU RT cores is already well defined, there are still research challenges to further tune the approach towards higher performance and energy efficiency (EE). In this work we introduce three ideas that can further improve particle FRNN physics simulations running on RT Cores; i) a real-time update/rebuild ratio optimizer for the bounding volume hierarchy (BVH) structure, ii) a new RT core use, with two variants, that eliminates the need of a neighbor list and iii) a technique that enables RT cores for FRNN with periodic boundary conditions (BC). Experimental evaluation using the Lennard-Jones FRNN interaction model as a case study shows that the proposed update/rebuild ratio optimizer is capable of adapting to the different dynamics that emerge during a simulation, leading to a RT core pipeline up to $\sim 3.4\times$ faster than with other known approaches to manage the BVH. In terms of simulation step performance, the proposed variants can significantly improve the speedup and EE of the base RT core idea; from $\sim1.3\times$ at small radius to $\sim2.0\times$ for log normal radius distributions. Furthermore, the proposed variants manage to simulate cases that would otherwise not fit in memory because of the use of neighbor lists, such as clusters of particles with log normal radius distribution. The proposed RT Core technique to support periodic BC is indeed effective as it does not introduce any significant penalty in performance. In terms of scaling, the proposed methods scale both their performance and EE across GPU generations. Throughout the experimental evaluation, we also identify the simulation cases were regular GPU computation should still be preferred, contributing to the understanding of the strengths and limitations of RT cores.
\end{abstract}

\begin{keyword}
Ray Tracing \sep RT Cores \sep GPU Computing \sep Nearest Neighbors \sep Particle Simulation \sep Fixed Radius

\end{keyword}

\end{frontmatter}

%-----------------------------------------------------------
\section{Introduction}\label{sec:introduction}

Particle simulation methods exist in a broad range of research fields, such as smoothed particle hydrodynamics (SPH), molecular dynamics (MD), and discrete element method (DEM)  \cite{monaghan1992smoothed,frenkel2002understanding,plimpton1995fast}. These fields share the fact that the computation of the interactions requires a fixed radius nearest neighbor (FRNN) search. In FRNN, for a system of $n$ particles, $p_1, p_2, \dots, p_n$,  one must identify all $\{p_i,p_j\}$ pairs in which their distance $r_{i,j}$ satisfies
\begin{equation}
    r_{i,j} = || p_i - p_j || < r_c
\end{equation} 
with $r_c$ being a fixed cutoff radius that is decided by the physics application. Algorithmically, the process can be viewed as finding all $p_j$ neighbors for each $p_i$. Finding the neighbors of each particle enables the computation of forces, densities, or collision responses depending on the physical model studied. Brute force solutions such as an all‑pairs search cost $O(n^2)$, introducing a significant performance cost for large simulations. For this reason, a significant amount of computational research has been put on improving the efficiency of the FRNN computation, specially when working with large‑scale simulations.

Efficient FRNN techniques normally rely on spatial data structures such as uniform grids, linked‑cell lists, kd‑trees, or octrees \cite{warren1993parallel,brown2011implementing}. These approaches, initially thought for CPUs, in the last years have transitioned to GPUs, bringing a significative performance uplift due to the GPU's highly parallel architecture \cite{navarro2014survey,amada2004particle,herault2010sph,krog2012fast}. Such GPU implementations are normally programmed with the general purpose programming model, offered by APIs such as CUDA or OpenCL. In such cases, the implementations take full advantage of the general purpose sections of the chip, such as the FP32/64 or INT units.

\begin{figure}[htbp]
  \centering
  \includegraphics[width=0.8\linewidth]{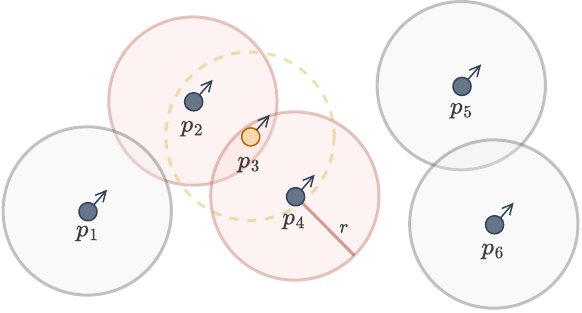}
  \caption{Fixed‑radius search with RT cores:
           a ray launched from particle $p_3$ intersects the bounding spheres of $p_2$ and $p_4$, which are its neighbors.}
  \label{fig:rt-nearest-neighbors}
\end{figure}

Recent hardware advances in GPU chip have introduced specialized compute units known as Ray Tracing (RT) cores, which sit next to the general purpose parts of the chip. Originally designed to accelerate real-time ray tracing for 3D rendering, RT cores can also be leveraged to handle the spatial geometric queries of the FRNN search. State of the art works~\cite{zhu2022rtnn,zhao2023leveraging,nagarajan2023rt} have demonstrated that by launching one ray per particle, RT cores can effectively capture the nearest neighbor particles within a given radius, as illustrated in Figure \ref{fig:rt-nearest-neighbors}.

While the fundamental concept underlying the utilization of RT cores in FRNN has been clearly established, further research remains necessary to fully explore its potential and limitations  towards further improving its performance and energy efficiency, as well as support boundary conditions and even optimizing the maintenance of the RT core's bounding volume hierarchy (BVH) along the simulation. In this context, \textbf{this work presents three contributions} that improve the performance and extend the usability of FRNN computations with RT cores for physics simulations:

\begin{itemize}
\item \textbf{Contribution \#1 - Optimal BVH rebuild/update ratio:} We propose \texttt{gradient}, an adaptive real-time technique that optimizes the ratio at which the BVH updates and rebuilds along the simulation. We recall that rebuilding the bounding volume hierarchy (BVH) of the RT core after each movement of particles would introduce a costly task at each time step. On the other hand, not rebuilding the BVH, and relying only on BVH "updates" (last level modifications on the tree) will progressively degrade the performance of each ray tracing query, to the point of being much worse than rebuilding. 

\item \textbf{Contribution \#2 - ORCS:} The base RT Core idea populates a neighbor list which is then used to compute forces and particle displacement with regular GPU compute kernels. The whole neighbor list can be of size $n\cdot k_{\text{max}}$, where $k_{\text{max}}$ is the maximum number of neighbors per particle, and $n$ the number of particles. In controlled small radius scenarios this is not an issue, as $k_{\text{max}}$ can be safely assumed low. However, clusters of particles with variable radius may exhibit a $k_{\text{max}} \sim n \implies \mathcal{O}(n^2)$ scenario, making the simulation run out of memory very soon in $n$. As a solution, we propose ORCS (Optimized Ray tracing Core Simulation) through two variants, \texttt{ORCS-pers\'e} and \texttt{ORCS-forces}, that do not need neighbor list, and are even faster for small and log normal radius.   

\item \textbf{Contribution \#3 - Ray traced periodic boundary conditions:} %When working with periodic boundary conditions, particles near the boundary of the simulation may be at a distance that is less than the radius search. For these particles, the fixed radius search must search on the space of the opposite boundaries. 
Handling different boundary conditions (BC) has not been a focus of research in the previous RT core works; because of this, the base RT core idea as it is works with wall BC but not for periodic BC. Here, we propose a new way to handle periodic BC with ray tracing, without the need to incur into regular GPU kernels for simulating the particles close to boundaries, neither replicating them, enabling RT Core FRNN for physics simulations with periodic boundary conditions. 
\end{itemize}

%These contributions will allow computational physics simulations based on FRNN to further improve their time and memory efficiency, enabling the possibility to tackle very large scale particle simulations.

The remainder of this manuscript is organized as follows: Section \ref{sec:related-work} discusses related work and Section \ref{sec:rtx-tuning} details the three proposed optimizations and their integration with RT cores. Section \ref{sec:experimental-evaluation} presents experimental results and performance analyses. Finally, Section \ref{sec:conclusions} provides discussion, conclusions and outlines future research directions.

\section{Related Work}
There is a considerable amount of research and recent results on CPU/GPU FRNN approaches, using both traditional GPU computing and leveraging RT Cores for neighbor-search. There is also work on performance optimizations to the BVH data structure. 
\label{sec:related-work}
\subsection{Traditional CPU and GPU FRNN Approaches}
Efficient nearest-neighbor search algorithms have long been a subject of research, particularly in the context of large-scale particle simulations and in the case of Fixed Radius Nearest Neighbors (FRNN). 
In FRNN, traditional CPU-based methods typically employ spatial decomposition structures such as grids (also known as cell-lists) \cite{ihmsen2011parallel,in2008accurate}, kd-trees \cite{foley2005kd,ram2019revisiting}, and octrees \cite{warren1993parallel,fernandez2022fast}. These data structures facilitate pruning of the search space, reducing the computational cost from $O(n^2)$ to $O(n\log n)$ or even less in the average case, depending on both the data distribution, radii and implementation details.

On GPUs, the uniform grid remains a reliable method for FRNN in many-particle systems \cite{goswami2010interactive,gpuCellList2013,chisholm2020improved}. Hierarchical GPU data structures have also been explored for FRNN, such as bounding volume hierarchies (BVHs) \cite{howard2019quantized}, kd-trees \cite{moreland2013comparing} and octrees \cite{zhou2010data}. While these methods provide high performance, handling rapidly moving particles introduces overheads in terms of tree rebuilds or updates \cite{brown2011implementing}.

\subsection{Ray Tracing Cores for Neighbor Search}
%Research in this category is the most relevant for this work, as it sets the starting ground from which the proposed contributions build from. Recent GPU ray tracing hardware, known as Ray Tracing Cores or simply RT Cores, opened the opportunity to research new ways to accelerate GPU Computing workloads \cite{meneses2024accelerating,geng2025librts,wald2019rtx,lv2024rtscan}, now by adapting them to the ray tracing pipeline which is hardware accelerated. 
%One of the most relevant uses of RT cores outside graphics is neighbor search. 
By formulating nearest-neighbor queries as intersections between rays and particles (i.e., particles surrounded by spheres or bounding volumes), one can detect nearest neighbors in parallel and accelerate this process by hardware-level BVH traversal logic from the GPU RT cores. Recent works have demonstrated the efficacy of this approach \cite{zhu2022rtnn,zhao2023leveraging,nagarajan2023rt,zellmann2020accelerating}, providing up to an order of magnitude of speedup over existing traditional GPU approaches. Conceptually, the technique is based on the idea of generating spheres of radius $r$ (the cutoff radius) in the location of each particle, and launching infinitesimally small rays in parallel, one for each particle, at the particle's positions (see Figure \ref{fig:rt-nearest-neighbors}). Then, the intersections between these rays and spheres (ignoring their own one) serve as a mechanism to discover the neighbors of each particle and register them in a neighbor list. One aspect that has not been fully studied yet is the management of periodic BC with RT cores. By default, the intrinsic locality of rays make the approach suitable for wall BC, but not for periodic BC. Another aspect that is relevant is the use of heuristics that could optimize the ratio of BVH updates/rebuild along the simulation which is a known challenge  \cite{burtscher2012quantitative}. 

\subsection{Improving the BVH for Dynamic Scenarios}
The cost of updating or rebuilding the RT core's BVH cannot be ignored, as this data structure was originally conceived in the context of static scenes \cite{wald2007ray}. Particle simulations are dynamic systems, thus the RT core's BVH must be kept up do date along the entire simulation. Performing BVH rebuilds after each simulation step ensures a balanced and optimal tree structure for all ray tracing queries, but at the cost of building time. On the other hand the BVH update (refitting), although a cheap operation that guarantees correctness, produces a unbalanced and non-optimal BVH tree that can progressively degrade the performance of all ray tracing queries \cite{meister2021survey}. Several works of the pre-RT core era have proposed software methods to mitigate this issue. Some have proposed faster building heuristics \cite{yin2014fast,pantaleoni2010hlbvh} based on improvements to the Surface Area Heuristic (SAH) \cite{havran2000heuristic}. Other works have managed to slow the degradation from the refit process of the BVH \cite{parker2010optix} or proposed new refitting approaches \cite{yin2014fast} and variants such as the hierarchical linear bounding volume hierarchy (HLBVH) \cite{viitanen2017mergetree,su2024quicktree}. The work by Lauterbach \textit{et al.} proposed an heuristic for improving the ratio between BVH refits and rebuilds \cite{lauterbach2006rt} based on the ratio between the surface area of the parent nodes and their children nodes combined with threshold value in the range $[0..100\%]$ that triggers a rebuild. This threshold must be set manually, and the authors found that $40\%$ did work efficiently for their benchmarks. Although this last work does find a way to improve the ratio between refit and rebuild when compared to other static approaches that rebuild after a fixed number of time steps, it requires access to the internal BVH tree structure which is not exposed in RT cores, and it also requires an optimal threshold value which may depend on the specific application. 

Currently, tuning FRNN based simulations with RT cores is a topic that has still has some unexplored research challenges. We identify three of them that can further improve RT Core FRNN in terms of performance and energy efficiency; i) finding a real-time RT core BVH rebuild/update ratio optimizer, ii) support periodic BC purely with RT core rays and iii) perform RT core FRNN simulations without the need for neighbor lists. In the following two sections, we propose new ideas for each of these challenges and show their performance and energy efficiency on simulations with representative particle/radius distributions using the Lennard-Jones (LJ) interaction model.

\section{Tuning the FRNN-based Simulations for RT Cores}
\label{sec:rtx-tuning}
We use the Lennard-Jones (LJ) potential \cite{jones1924determination,allen2017computer} as FRNN simulation case study for the remainder of this work, as it is a widely known interaction model for molecular dynamics and computational physical chemistry, among other fields. For a pair of particles $i,j$, the LJ potential is defined as 
\begin{equation}
    U_{ij}(r) = 4\epsilon \left[\left(\frac{\sigma}{r}\right)^{12} - \left(\frac{\sigma}{r}\right)^{6}\right]
\end{equation}
Where $\epsilon$ is a strength factor known as the depth of the potential well, $r$ is the distance between the particles $i$ and $j$ from their centers, $\left(\frac{\sigma}{r}\right)^{12}$ is the attraction term, $\left(\frac{\sigma}{r}\right)^{6}$ is the repulsion term, and $\sigma$ defines the distance at which the potential is zero. In practical cases, long-range interactions with the LJ potential are almost negligible in comparison to the near-range ones. For this, LJ is usually implemented using a cutoff radius, here denoted as $r_c$, such that
\begin{equation}
\label{eq:LF-cutoff}
        U_{ij}(r) =  
        \begin{cases}
        4\epsilon \left[\left(\frac{\sigma}{r}\right)^{12} - \left(\frac{\sigma}{r}\right)^{6}\right], & r \le r_c\\
        0,              & r > r_c
        \end{cases}
\end{equation}
This reduces the amount of necessary interactions per time step in the system, from $\Omega(n^2)$, to a bound of $O(n \cdot k)$ where $k$ would be the maximum number of nearest neighbors found for one particle. In practice, simulations move their particles by computing the forces between them, which are defined as the negative gradient of $U$,
\begin{equation}
\label{eq:LF-forces}
        F_{ij}(r) = -\nabla{U}_{ij}(r) =
        \begin{cases}
        24\epsilon \left[\left(\frac{\sigma}{r}\right)^{12} - \left(\frac{\sigma}{r}\right)^{6}\right]\frac{1}{r}, & r \le r_c\\
        0,              & r > r_c
        \end{cases}
\end{equation}
Equations \ref{eq:LF-cutoff} and \ref{eq:LF-forces} have assumed a model where $\sigma$ and $r_c$ are common to all particles, but they can also be set per-particle, leading to a case of variable radius. In LJ simulations, usually $k \ll n$, but extreme situations with a particular combination of particles and radius distributions (such as a dense cluster) might imply a much larger $k$ value, or even $k \sim n$. In such cases, the initial simulation steps have very intense interactions, and over time the system stabilizes thanks to the repulsion term of Eq. (\ref{eq:LF-cutoff}). Such cases have also been considered in the experimental evaluation of Section \ref{sec:experimental-evaluation}.
%As mentioned in Section \ref{sec:related-work}, state-of-the-art GPU-based FRNN methods 
%for computing the LJ forces from Eq. \ref{eq:LF-forces}  range from cell-list (a.k.a grid) based methods to space partition trees (octrees, kd-trees, BVH) and recently the use of hardware-accelerated ray tracing with RT cores (Figure \ref{fig:rt-nearest-neighbors}). 
In the next subsections we present the three improvements proposed for FRNN with RT cores.

\subsection{Gradient: a real-time update/rebuild ratio optimizer}
\label{subsec:bvh-rebuild-update}
\begin{figure}[ht!]
    \centering
    \includegraphics[width=0.98\linewidth]{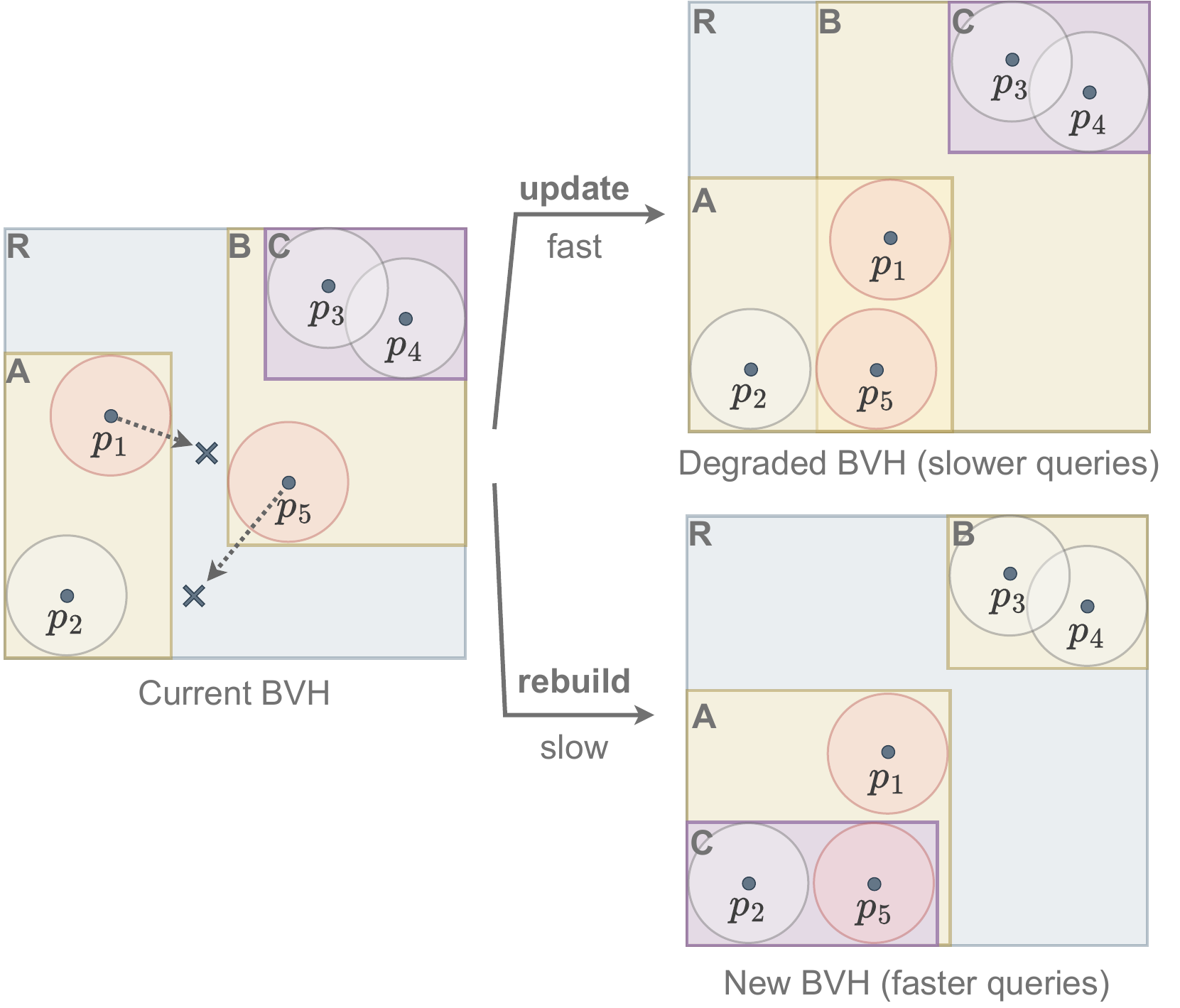}
    \caption{An example simulation of five particles. The left side shows the current state of its BVH and the simulation of one time step which moves particles $p_1$ and $p_5$. On the right, the resulting BVH by applying update or rebuild.}
    \label{fig:bvh-build-update}
\end{figure}

In RT core programming, there are two functions that allow the programmer to manage the intenral BVH, i) \texttt{build} and ii) \texttt{update}. The \texttt{build} function receives a set of primitives in space as input (among other configuration parameters), and builds the corresponding hardware-accelerated BVH, which is optimized for that current spatial distribution of primitives. In FRNN, these primitives are the spheres that represent the particles and their search radius (Figure \ref{fig:rt-nearest-neighbors}). On the other hand, the \texttt{update} function updates an existing BVH to the current state of positions of the primitives by applying minimal changes to the BVH. Currently, these minimal changes are achieved through \text{refitting}, which consists of expanding (or shrinking) the bounding volumes of the last level nodes to adapt to the new positions of the primitives. Figure \ref{fig:bvh-build-update} illustrates an existing set of primitives with its BVH structure, and the effects of updating vs rebuilding after a simulation step that moved the particles. 

\begin{figure}[ht!]
    \centering
    \includegraphics[width=1.0\linewidth]{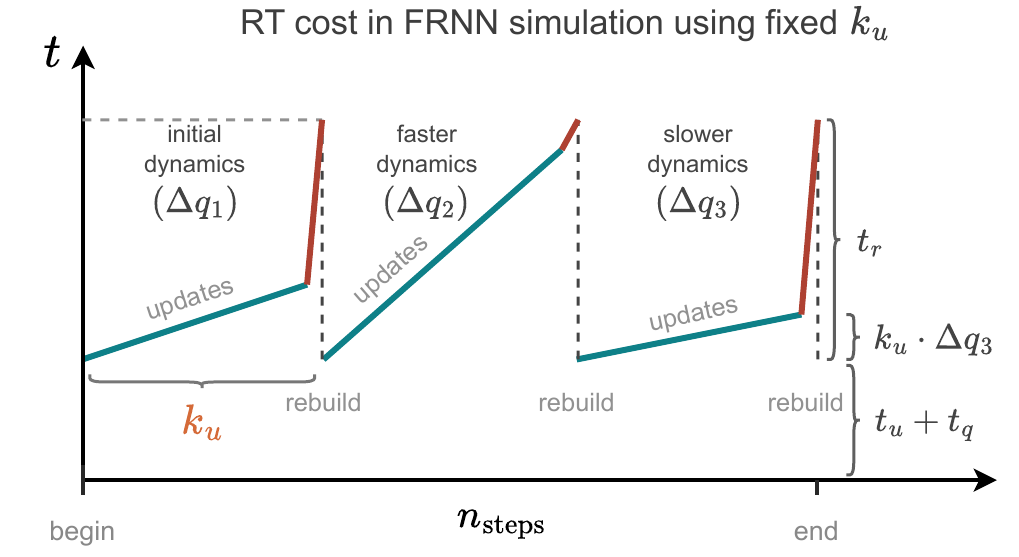}
    \caption{Cost curve using a fixed number of updates $k_u$ on the BVH.}
    \label{fig:rtxnn-build-update-curve}
\end{figure}

Although the \texttt{update} procedure is a fast process, as particles move because of the physical forces computed from the FRNN interactions, it progressively makes the BVH slower at RT queries as each ray intersects more overlapping bounding boxes. On the other hand, \texttt{rebuild} is slow as it builds the entire BVH from scratch, but in return the BVH is optimized for the new spatial state of particles. With moving particles, the challenge is to find the optimal number of consecutive updates, namely $k_u$, before doing a full rebuild. This can be seen as optimizing the area under the curve of the plot RT time ($t$) vs simulation steps ($n_{\text{steps}}$). Figure \ref{fig:rtxnn-build-update-curve} illustrates an example using a fixed $k_u$ for a simulation that undergoes different types of dynamics, from faster to slower.

The simulation cost curve from Figure \ref{fig:rtxnn-build-update-curve} is composed of the following parameters:
\begin{itemize}
    \item $n_{\text{steps}}$: number of simulation steps.
    \item $t_r$: BVH rebuild cost.
    \item $t_u$: BVH update cost.
    \item $t_q$: RT query cost with a freshly built BVH.
    \item $\Delta q$: average extra RT query cost for using updated BVH.
    \item $k_u$: number of consecutive steps with BVH update.
\end{itemize}
Using a constant $k_u$, such as in Figure \ref{fig:rtxnn-build-update-curve}, where the BVH is just updated for $k_u$ consecutive steps, followed by a full rebuild, is not the best choice in dynamic cases as it does not take into account the dynamics of the simulation; i.e., faster dynamics would benefit from a lower value of $k_u$ while slower dynamics would benefit from a higher value of $k_u$. Not adjusting $k_u$ potentially leads to simulation steps with too much BVH degradation, or rebuilds that occur too early on BVHs that still have not degraded enough. As a solution we propose \texttt{gradient}; an adaptive real-time update/rebuild ratio optimizer. This is achieved by formulating the following cost model: 
\begin{equation}
    T_{\text{sim}} = \frac{n_{\text{steps}}}{k_u+1}\left[\left(\frac{k_u \cdot (k_u \Delta q)}{2}\right) + k_u(t_u + t_q) + (t_r + t_q)\right]
\end{equation}
which computes the total RT cost of the simulation as the area under the curve illustrated in Figure \ref{fig:rtxnn-build-update-curve}. Taking its derivative equal to zero leads to the following quadratic equation
\begin{align}
\frac{\partial T_{\text{sim}}}{\partial k_u} &= 0\\
\Delta q k_u^2 + 2\Delta q k_u + 2(t_u - t_r) &= 0
\end{align}
where its solution suggest an optimal $k_u$ of
\begin{equation}
    k_u^{\text{opt}} = -1 + \sqrt{1 - 2\left( \frac{t_u - t_r}{\Delta q}\right )}.
\end{equation}
To make it adaptive, the value $k_u^{\text{opt}}$ is kept updated through a time window, by keeping track of $t_u, t_r$ and $\Delta q$ with time measurements on each step using NVIDIA NVML (GPU) \cite{nvidia_nvml}. 

\subsection{ORCS: RT core FRNN simulation with no neighbor list}
\begin{figure*}
    \centering
    \includegraphics[width=\linewidth]{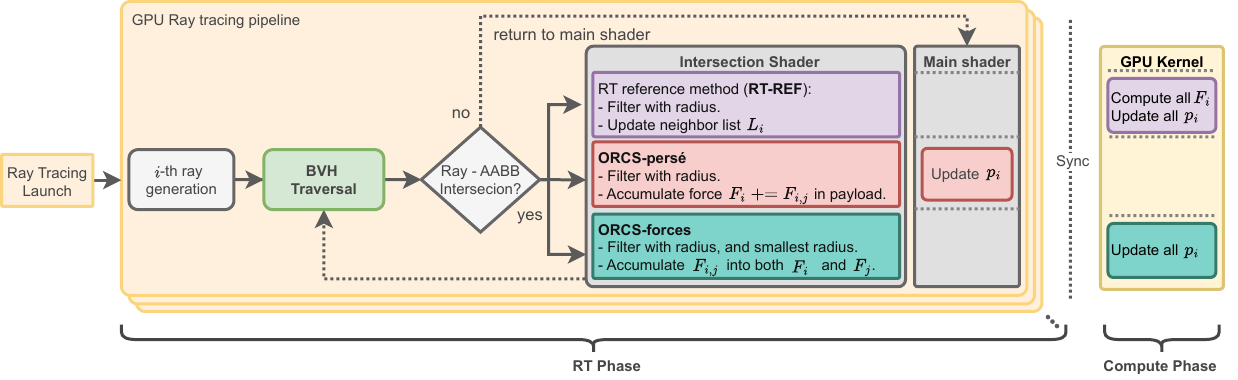}
    \caption{The pipeline for one simulation step using RT cores. The purple boxes represent the traditional way of leveraging RT Cores, while the red and teal ones are the proposed ORCS variants with no need for neighbor lists.}
    \label{fig:rtx-physics}
\end{figure*}

The original idea of using RT Cores for FRNN simulations is to accelerate the neighbor finding phase. Normally, in existing RT core approaches, the output is a neighbor list, which is then given as input to a separate GPU kernel where all forces $F_{i,j}$ are accumulated into each $F_i$, and then applied in parallel to the particles $p_i$. This is illustrated in Figure \ref{fig:rtx-physics} with the pipeline following the top purple variant. 
We present \textbf{ORCS} (Optimized Ray tracing Core Simulation) through two variants that can work without neighbor lists; i) \texttt{ORCS-pers\'e} and ii) \texttt{ORCS-forces}. 

\subsubsection{ORCS-pers\'e}
\label{subsubsec:ORCS-perse}
In this variant the $i$-th ray tracing thread simulates $p_i$. It begins by initializing an empty force vector in a payload structure, and launches its corresponding ray with the payload attached. The payload travels along with the ray through the intersections, and each time the ray hits an axis aligned bounding box (AABB), the intersection shader is triggered, and if the particle is in fact inside the search radius, then a new force contribution is computed and independently accumulated into the force vector that resides in the payload. After the ray finishes with all intersections, the payload's final force vector is used to compute and update the particle's position $p_i$, which is stored in GPU global memory. This variant does not require any neighbor list or additional compute kernels, although is limited to simulations where all particles have the same radius.

\subsubsection{ORCS-forces}
\label{subsubsec:ORCS-forces}
\begin{figure}[ht!]
    \centering
    \includegraphics[width=0.4\linewidth]{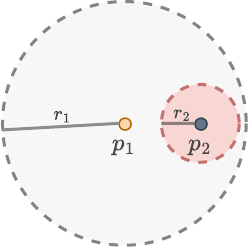}
    \caption{Variable radius case: only $p_2$'s ray triggers an intersection, therefore such thread is in charge of adding the force contributions on both $p_1$ and $p_2$.}
    \label{fig:rtx-variable-radius}
\end{figure}

Here, instead of using the payload, each RT intersection shader computes the interaction force $F_{i,j}$ and accumulates it into both $F_i$ and $F_j$, which reside in GPU global memory. This variant can work in both uniform and variable radius. In the uniform radius scheme, whenever an intersection shader triggers, the thread with smaller \texttt{id} will compute $F_{i,j}$ and apply it on the $F_i$ and $F_j$ forces located in GPU global memory, atomically. In the variable radius scheme, the interaction is not necessarily symmetric, therefore the thread with the smallest search radius propagates $F_{i,j}$ to both particles atomically. Figure \ref{fig:rtx-variable-radius} illustrates a non-symmetric case where particle $p_2$'s ray is the only one that can provide the force contributions to both $p_1$ and $p_2$.

\subsubsection{Differences between pers\'e and forces variants}
Although the two variants are capable of simulating particles without needing a neighbor list in GPU memory, there are still differences to highlight. The main one is that the \texttt{pers\'e} variant can do the whole simulation step within RT without requiring an extra GPU computing kernel although it is restricted to simulations with the same radius across all particles. On the other hand, \texttt{forces} computes all the forces $F_i$ in RT and applies them in a separate compute kernel. It supports both uniform and variable radius but requires atomic operations which could impose a potential performance penalty. 

\subsection{Ray traced periodic boundary conditions}
\begin{figure}[ht!]
    \centering
    \includegraphics[width=1.0\linewidth]{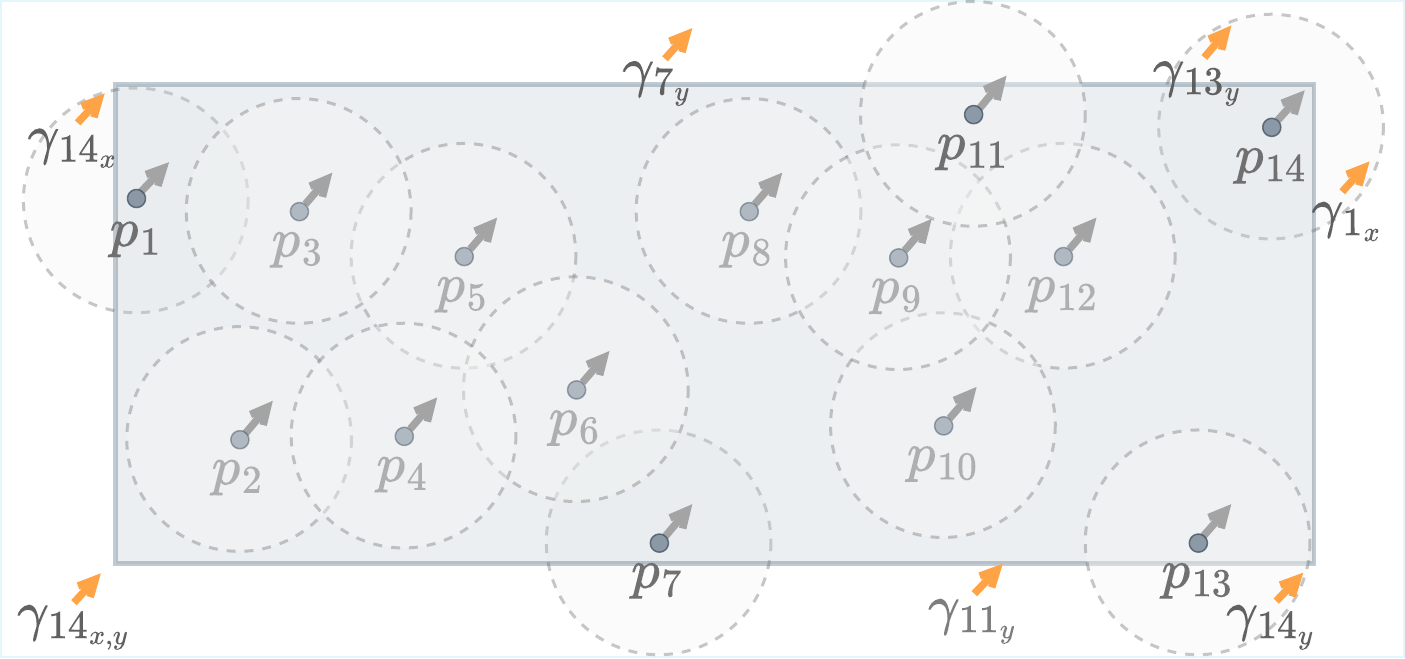}
    \caption{RT core scheme for periodic boundary conditions. Particles $\{p_1, p_7, p_{11}, p_{13}, p_{14}\}$ launch extra rays $\{\gamma_1,\gamma_7,\gamma_{11},\gamma_{13},\gamma_{14_x},\gamma_{14_y},\gamma_{14_{x,y}}\}$.}
    \label{fig:rt-boundary-bc}
\end{figure}

Periodic boundary conditions is a case of interest for particle simulations. In a RT scheme, any ray launched on a particle at the boundary of the simulation box will not detect neighboring particles located on the opposite side of the box. One way to handle this issue within ray tracing is to launch an extra ray $\gamma_i$ for each boundary wall where $p_i$'s search radius has passed through. Figure \ref{fig:rt-boundary-bc} illustrates an example where $\gamma$ rays are launched for particles $p_1, p_7, p_{11}, p_{13}, p_{14}$, with $p_{14}$ launching three extra rays $\gamma_{14_x}, \gamma_{14_y}$ and $\gamma_{14_{x,y}}$.

The main advantage of this approach is that it does not require creating extra geometry, as all is handled with rays. In the case of constant uniform radius, this idea is more efficient, but in the case of variable radius, the approach must trigger the gamma rays when the distance between $p_i$ and a boundary is smaller than the largest radius in the whole simulation, in order to cover the case similar to Figure \ref{fig:rtx-variable-radius} but when the larger radius particle is on the opposite side of the simulation box. This method of handling variable radii has a potential worst case where just one particle has a very large search radius, forcing all other particles to launch gamma rays if their distance to the boundary is within that search radius. Experimental results in Section \ref{sec:experimental-evaluation} show the behavior of this approach for both uniform and variable radii.

\section{Experimental Evaluation}
\label{sec:experimental-evaluation}
The implementations of the proposed ideas as well as the reference ones have been made available at \url{https://github.com/temporal-hpc/ORCS}\footnote{The repository will be accessible with the published article.}, where \texttt{ORCS} stands for "Optimized Ray tracing Core Simulation". Benchmarks use the Lennard Jones (LJ) model as case study, simulating particles in a 3D box of size $10^3 \times 10^3 \times 10^3$ using different combinations of particle and radius distributions. The entire experimental evaluation consists of three benchmarks:
\begin{itemize}
    \item \textbf{BVH rebuild/update schemes:} This benchmark evaluates the performance of \texttt{gradient} (the proposed BVH update/rebuild ratio optimizer presented in Section \ref{subsec:bvh-rebuild-update}), and compares it to other known policies such as rebuilding at a fixed rate or just rebuilding once the average cost of RT with updates surpasses the average cost of an RT step with rebuild.
    \item \textbf{Simulation performance:} This benchmark evaluates the average time and speedup of the proposed approaches \texttt{ORCS-pers\'e} and \texttt{ORCS-forces} and compares them with the base RT core approach (\texttt{RT-REF}) as well as with state-of-the-art parallel CPU/GPU approaches that employ the cell list method. Tests cover both wall and periodic BC.
    \item \textbf{Energy efficiency and scaling:} This benchmark measures power consumption, total energy and energy efficiency (interactions per Joule) for all approaches. We also include results showing how the raw performance and energy efficiency scale across different GPU generations, revealing a possible trend for future GPUs. 
\end{itemize}
Experimentation was conducted on the Patag\'on Supercomputer of Austral University of Chile \cite{patagon}, with the following hardware:
\begin{table}[ht!]
    \centering
    \resizebox{\columnwidth}{!}{
        \begin{tabular}{|l|l|}
        \hline
        Hardware  &  Specifications \\
        \hline
        OS        &  Ubuntu 24.04 LTS 64-bit\\
        CUDA      &  CUDA v13.0\\
        CPU       &  AMD EPYC 9534 64-Core Processor\\
        RAM       &  768 GB DDR5\\
        \textbf{GPU}       &  \textbf{NVIDIA RTX Pro 6000 Blackwell Server Edition (RTXPRO)}\\
        \hline
        \end{tabular}
    }
    \caption{Hardware used for the experimental evaluation.}
    \label{tab:placeholder}
\end{table}

\subsection{BVH rebuild/update schemes}
The first benchmark consists of measuring the performance of the proposed BVH rebuild/update scheme presented in Section \ref{subsec:bvh-rebuild-update}, named \texttt{gradient}, and compare it to two other commonly known approaches : in \texttt{fixed-200} we rebuild the BVH each 200 time steps, and in \texttt{avg} we rebuild only once the average cost (since the last rebuild) of simulating without rebuilding surpasses the average cost of simulating with rebuild. All three policies were tested with the base RT core idea (not the variants), with simulations of $n=140k$ particles for 2000 time steps with periodic BC. Each test measures the RT cost of updating/rebuilding the BVH followed by the RT query, on different combinations of initial position and radius distributions. 

\begin{figure}[ht!]
    \centering
    \includegraphics[width=0.95\linewidth]{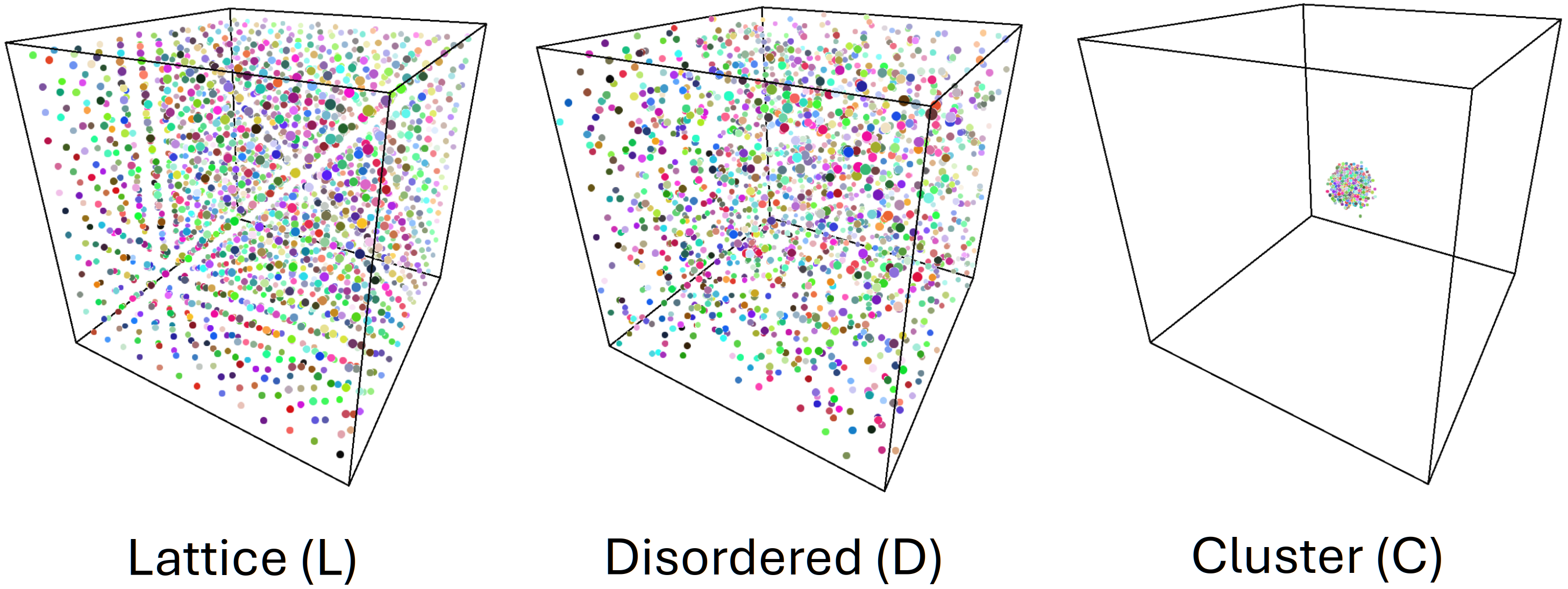}
    \caption{The three initial particle distributions used for the experimental evaluation. Each one is simulated under four different radius distributions.}
    \label{fig:particle-dists}
\end{figure}

\begin{figure*}[ht!]
    \centering
    \includegraphics[width=0.245\linewidth]{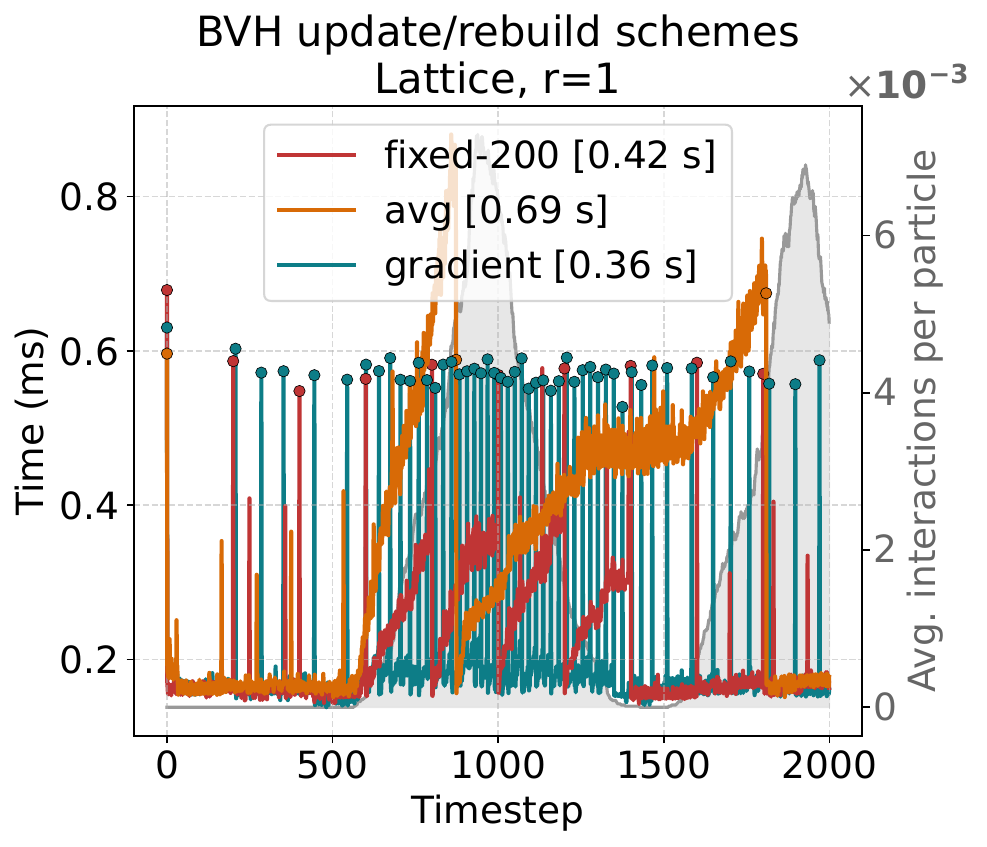}
    \includegraphics[width=0.245\linewidth]{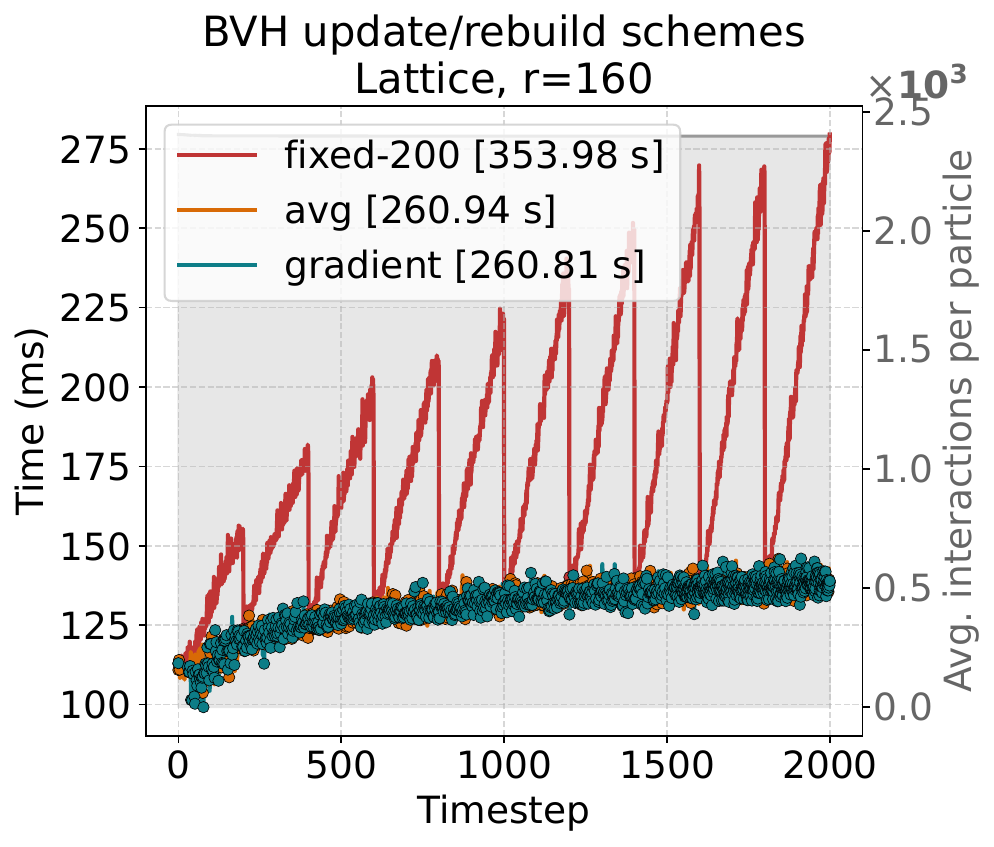}
    \includegraphics[width=0.245\linewidth]{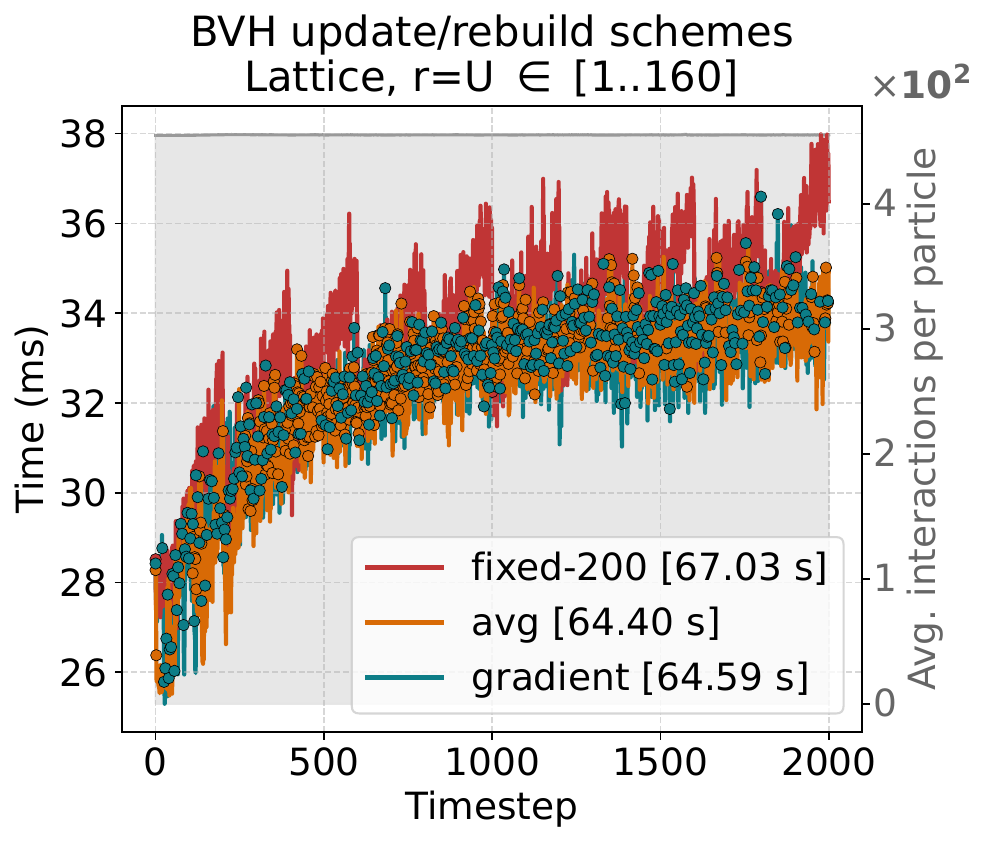}
    \includegraphics[width=0.245\linewidth]{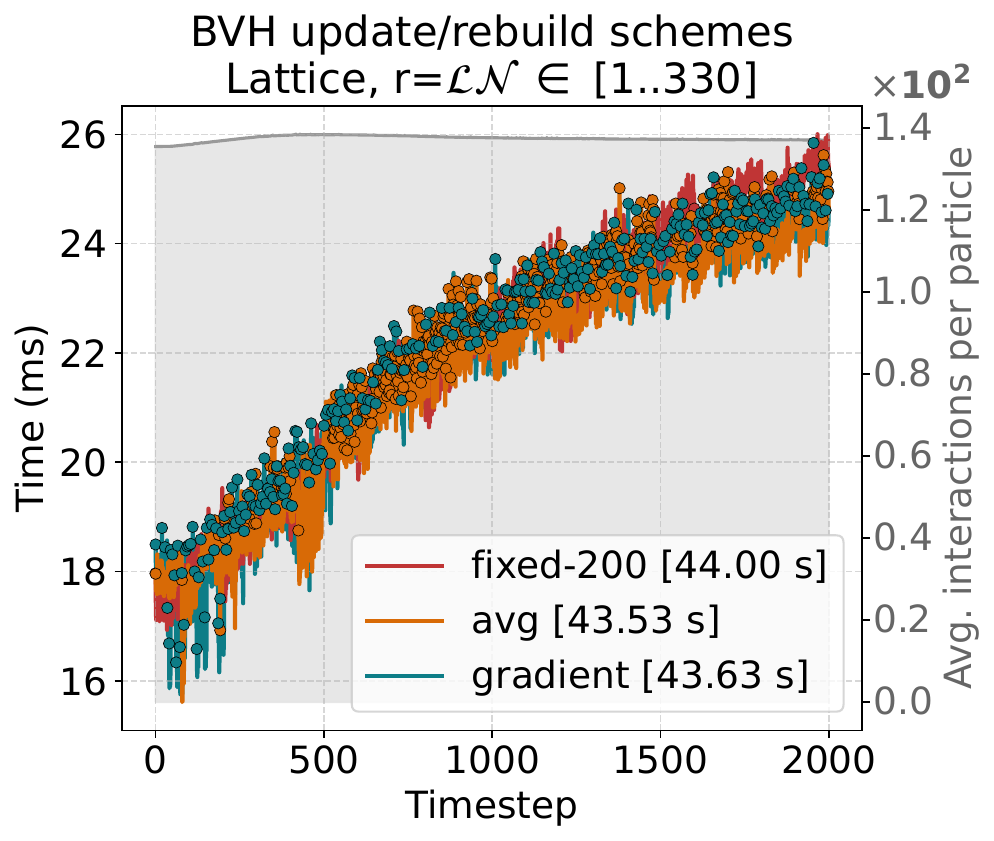}\\
    
    \includegraphics[width=0.245\linewidth]{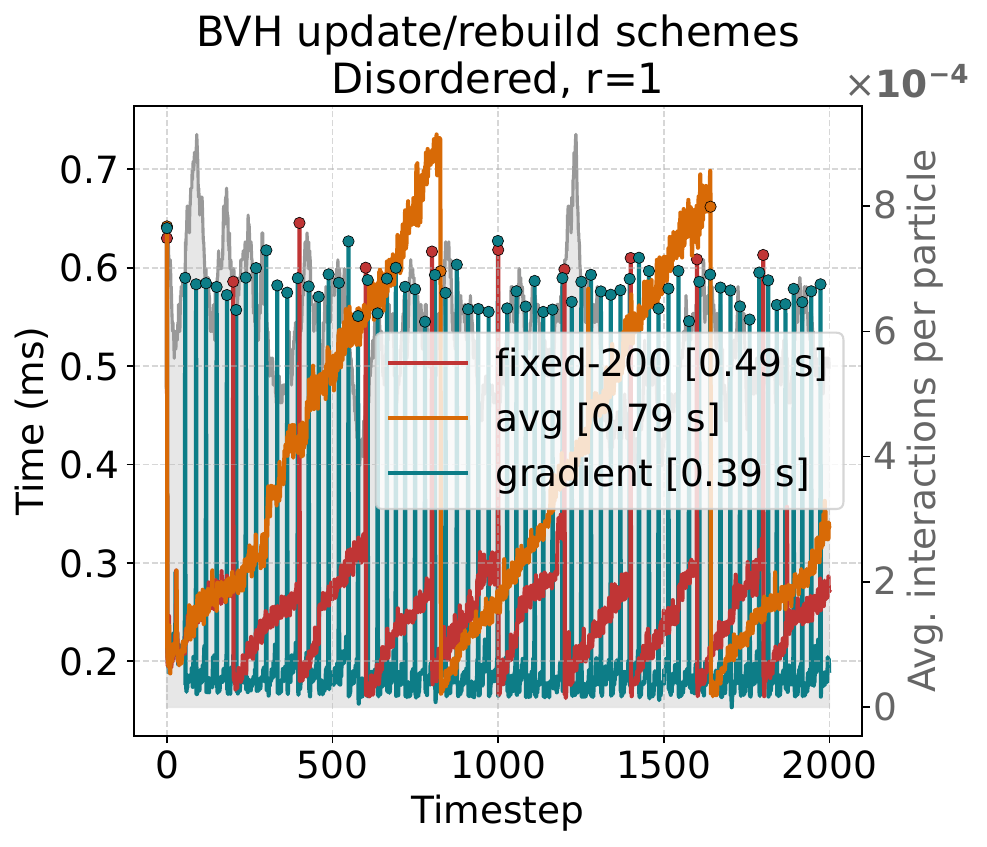}
    \includegraphics[width=0.245\linewidth]{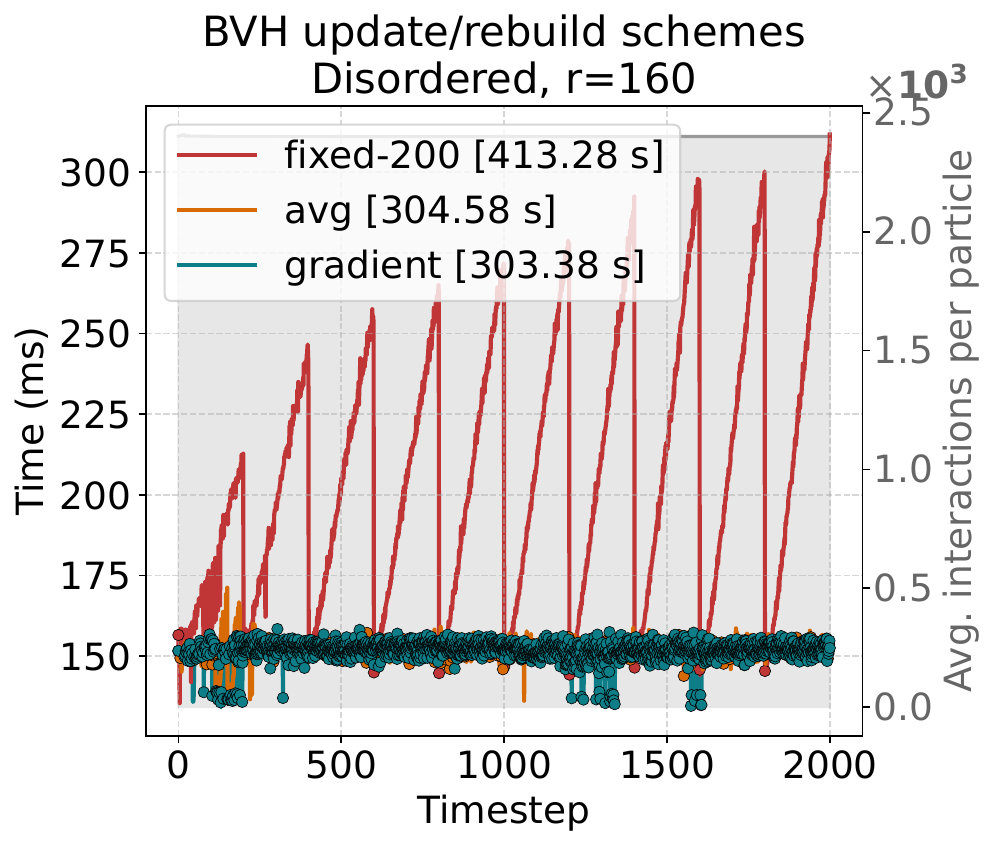}
    \includegraphics[width=0.245\linewidth]{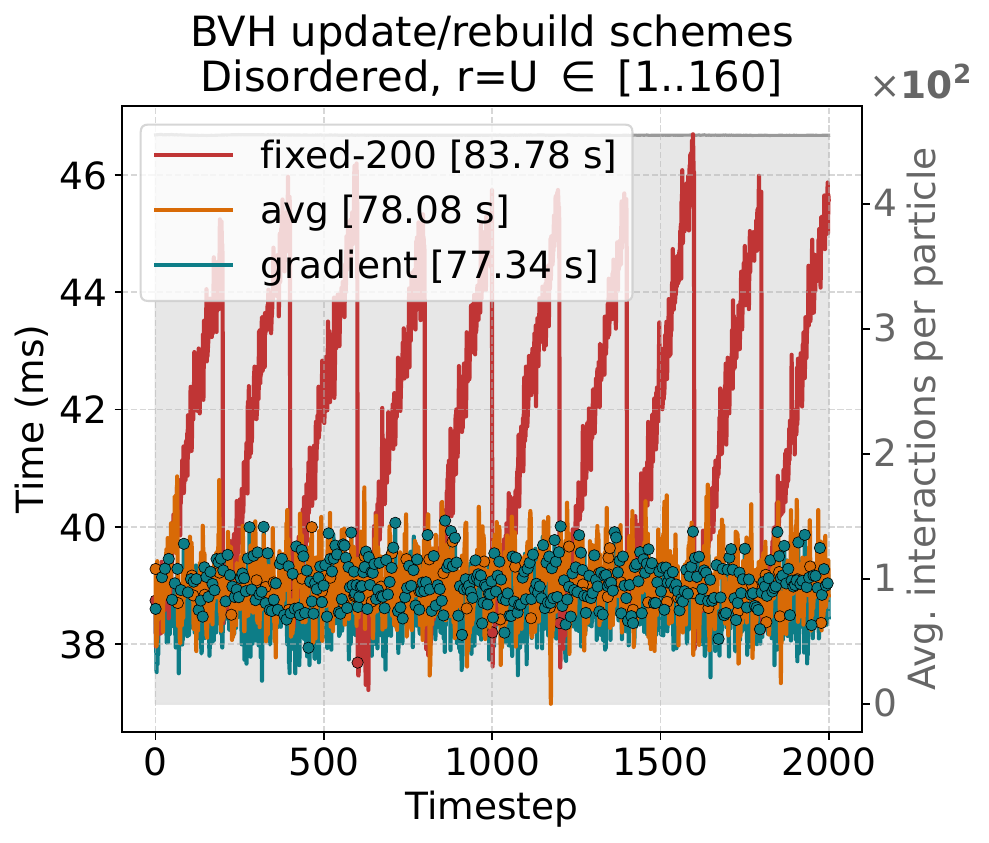}
    \includegraphics[width=0.245\linewidth]{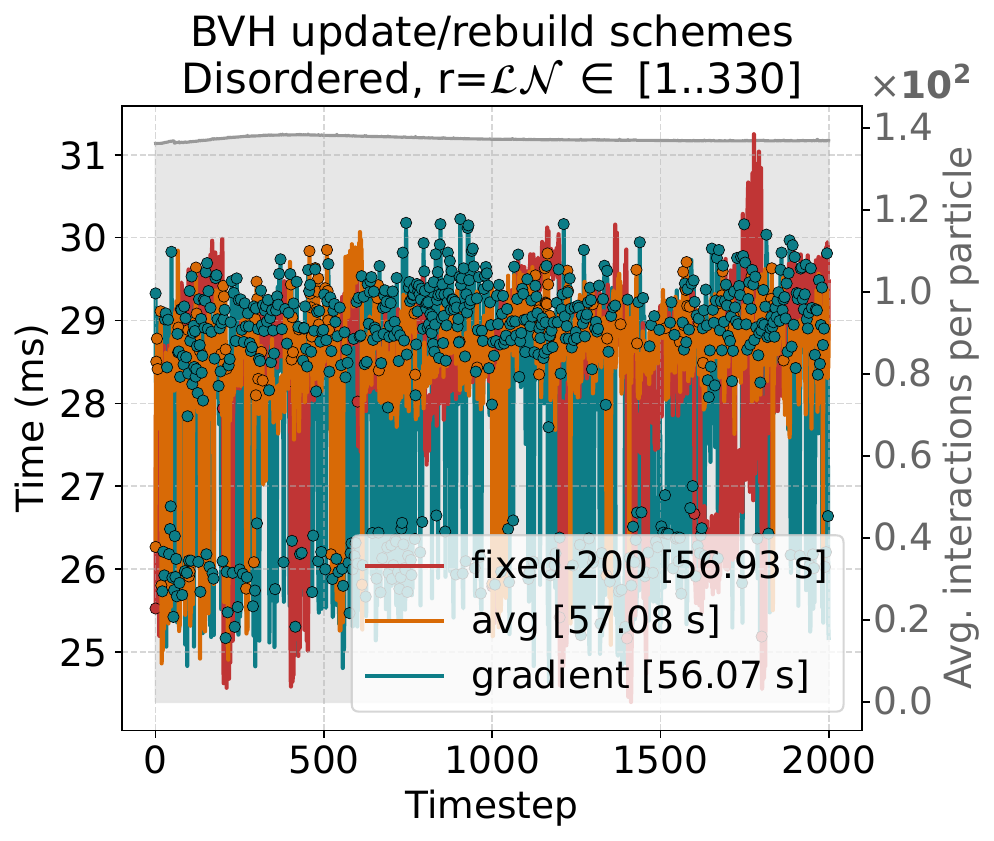}\\
    
    \includegraphics[width=0.245\linewidth]{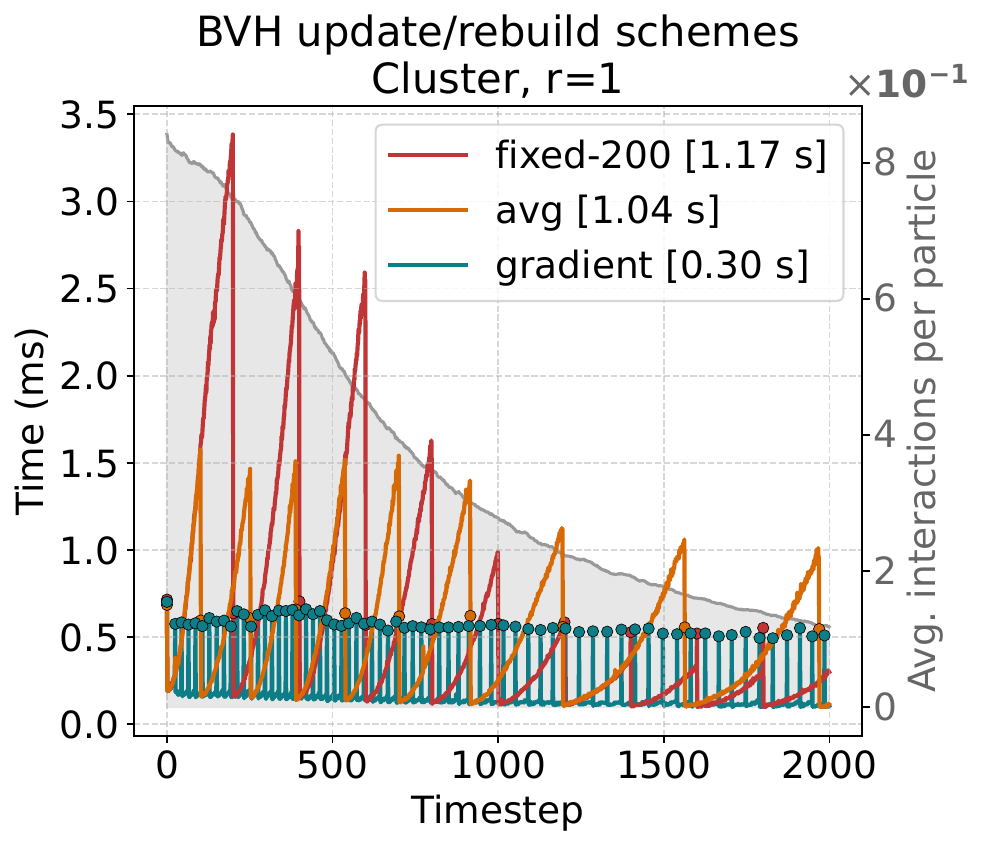}
    \includegraphics[width=0.245\linewidth]{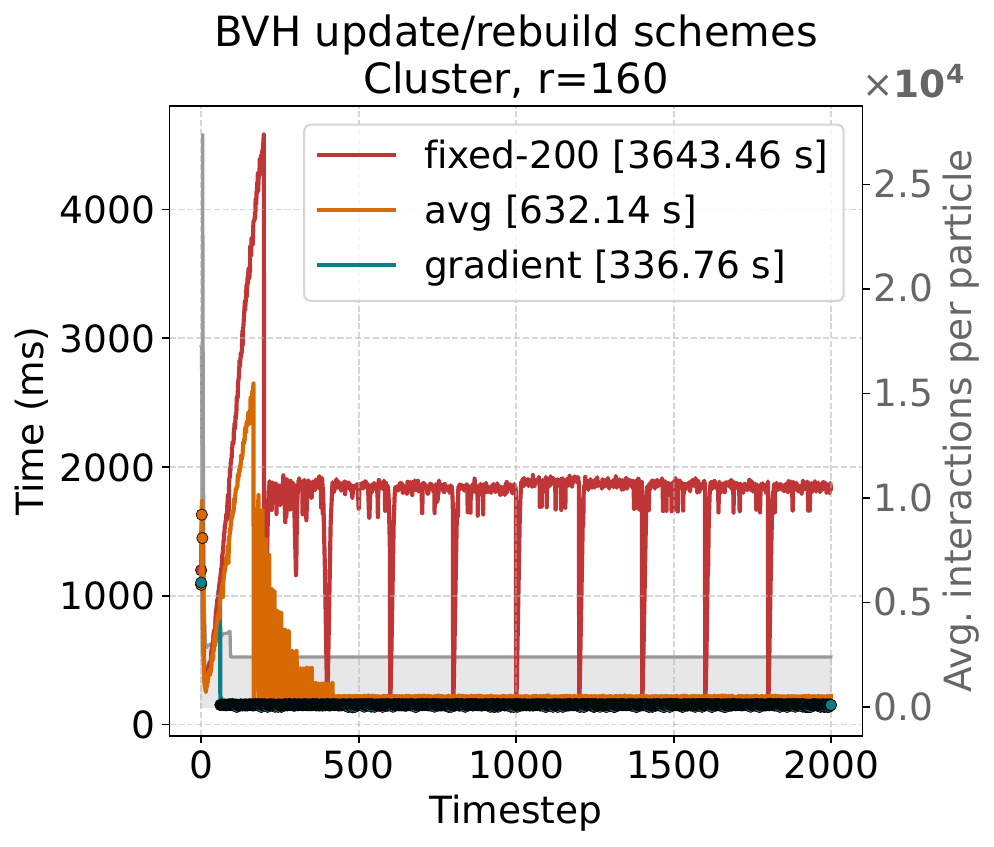}
    \includegraphics[width=0.245\linewidth]{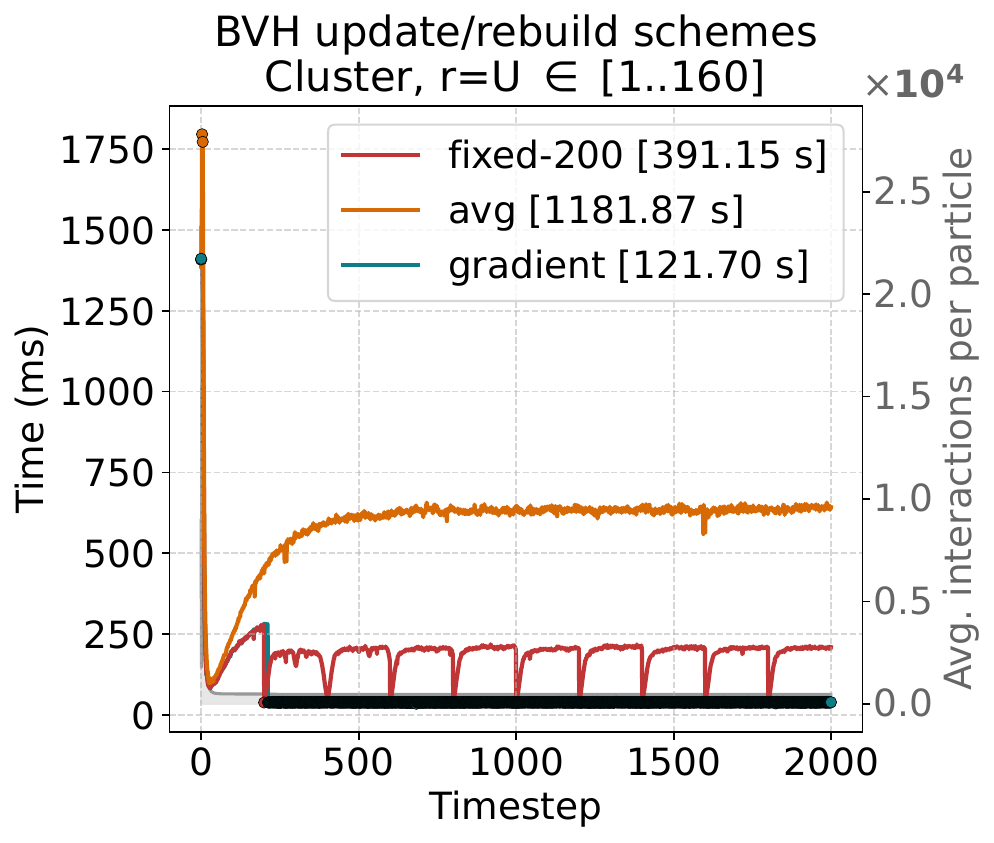}
    \includegraphics[width=0.245\linewidth]{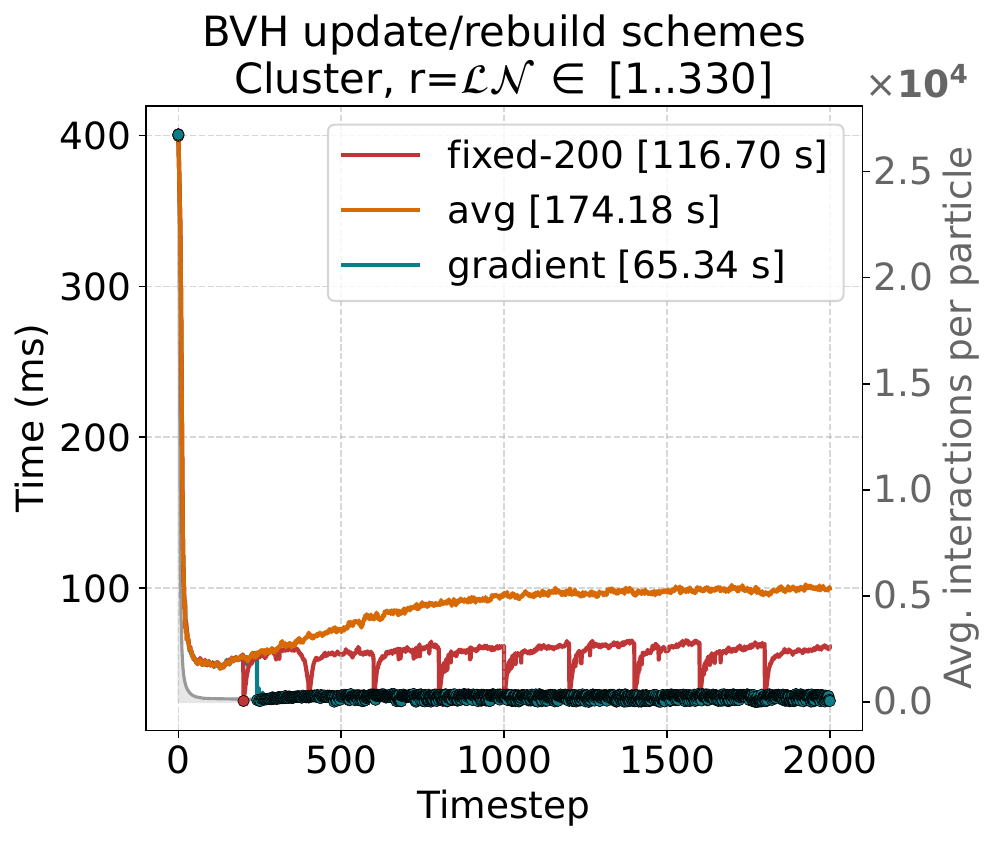}
    \caption{Time series of ray tracing cost for a simulation of $n=140k$ particles for 2000 time steps using periodic boundary conditions, under different particle (rows) and radius (columns) distributions.}
    \label{fig:bvh-schemes-results}
\end{figure*}

Three different initial particle distributions are tested; Lattice (L) through grid positions, Disordered (D) through a random uniform distribution and Cluster (C) through a random normal distribution $\mathcal{N}(\mu=\text{rand}, \sigma=25)$ as shown on Figure \ref{fig:particle-dists}. As for the radius, four different radius distributions are tested; small constant radius ($r=1$), large constant radius ($r=160$), random uniform distribution in the range $[1,160]$ and random log-normal distribution $\mathcal{LN}(\mu=1, \sigma=2)$ in the range $[1, 330]$.
Figure \ref{fig:bvh-schemes-results} presents the results for the three different schemes; the circle marks on each curve correspond to the moment when a rebuild is called. Also, the total cumulative time is in the legend of each curve and the average number of interactions per particle is in the second Y-axis \footnote{The average interactions per particle is a reference, but a high value does not necessarily mean high BVH degradation.} as a curve with its area shared in gray.  

The plots of Figure \ref{fig:bvh-schemes-results} show that \texttt{gradient} adapts to the dynamics of the simulation; rebuilding at a faster rate when the dynamics are faster, and at a slower rate when the dynamics are slower, resembling the scenario illustrated in Figure \ref{fig:rtxnn-build-update-curve}. The Lattice $r=1$ configuration is a clear example, as \texttt{gradient} manages to adapt to the oscillating dynamics, supported by the average number of interactions per particle, while \texttt{fixed-200} cannot adapt, and \texttt{avg} performs even worse as it is unable to react and rebuild fast enough due to the delay coming from the cumulative average since its last rebuild. In general, when the interaction radius is constant and small ($r=1$), \texttt{gradient} is between $1.16\times$ and $\sim 3.4\times$ faster than the second fastest approach. The second column of plots shows again that \texttt{gradient} is the fastest approach and also reveals that \texttt{fixed-200} can be the slowest approach when the dynamics are more chaotic, such as when all particles have large radii. In particular, in Cluster $r=160$, \texttt{avg} takes some time to adapt while \texttt{fixed-200} is penalized due to its fixed scheme, making \texttt{gradient} up to $1.87\times$ and $\sim 10.8\times$ faster than \texttt{avg} and \texttt{fixed-200}, respectively. The third column of plots uses a random uniform distribution of radius in the range $r \in [1,160]$. Here, in the Lattice and Disordered cases \texttt{gradient} is competitive with the other schemes, except for Cluster where it is $3.21\times$ faster than the second fastest approach (\texttt{fixed-200}). The fourth column tests the log normal radius distribution, where most of the particles have small radius except for a few. The performance behavior is similar to the third column, except for the relative increasing base cost in Lattice due to the later configurations being more complex than the initial one (grid). In the Cluster case, \texttt{gradient} is up to $1.8\times$ faster than the two other approaches.

\subsection{Simulation Performance}
The following approaches\footnote{RT-REF, ORCS-pers\'e and ORCS-forces use the \texttt{gradient} optimizer.} have been implemented and tested.
\begin{itemize}
    \item \textbf{CPU-CELL:} Reference parallel OpenMP CPU-based approach of the cell list approach \cite{ihmsen2011parallel}, adapted to support large dense scenarios by computing the forces array directly from the cell grid exploration.
    
    \item \textbf{GPU-CELL:} Reference GPU-based approach of the cell list algorithm. Its implementation builds from the one proposed by Crespin \textit{et al.} \cite{crespin2025soma} and further optimizes z-ordering with an out-of-place GPU radix-sort for Z-ordering, and removes the need of a fixed-size neighbor list, allowing dense cases to be simulated. 
    
    \item \textbf{RT-REF:} Reference method for FRNN with RT Cores \cite{zhu2022rtnn,zhao2023leveraging,nagarajan2023rt,zellmann2020accelerating}. Through ray tracing, it fills a neighbor list for later use in the physical simulation kernel. 
    
    \item \textbf{ORCS-pers\'e:} Proposed in sub-subsection \ref{subsubsec:ORCS-perse}, which does not require a neighbor list and does the whole physical simulation in the OptiX ray tracing pipeline, but with the restriction that all particles must have equal radius.
    
    \item \textbf{ORCS-forces:} Proposed in sub-subsection \ref{subsubsec:ORCS-forces}, which does not require a neighbor list, computes the forces array in the OptiX ray tracing pipeline, requiring an additional CUDA kernel, but compensates by supporting variable radius across particles.
\end{itemize}
Table \ref{tab:rtxpro_time} presents the running time of all five approaches under the different particle and radius distribution configurations, both under wall and periodic boundary conditions, and for small ($n=50k$) and large ($n=1M$) number of particles. In each configuration, the fastest approach is highlighted with teal color. For \texttt{ORCS-pers\'e}, we recall that it only works for constant and equal radius for all particles, while for \texttt{RT-REF}, its cells with no values correspond to out-of-memory cases because of its required neighbor list. The table shows that at low radius ($r=1$), \texttt{ORCS-pers\'e} is the fastest approach in almost all cases, followed by \texttt{RT-REF}. In the fixed large radius case ($r=160$) the scenario changes and the fastest approaches are the CELL based ones; \texttt{CPU-CELL@64c} or \texttt{GPU-CELL}, followed by \texttt{RT-REF} in the Lattice distribution. In the variable radius $r=U \in [1..160]$ case, the scenario is mixed as in the Lattice and Disordered distributions the fastest approaches are \texttt{RT-REF} and \texttt{ORCS-forces}, but in the Cluster distribution the CELL based ones dominate. Lastly, in the LogNormal radius distribution $r=\mathcal{LN} \in [1, 330]$, we find that \texttt{ORCS-forces} is the fastest by a significant margin, specially for a large number of particles. 

\begin{table*}[ht!]
\centering
\caption{Average time (ms) per step under different particle/radius and BC configurations. Cells marked in teal color are the fastest on each configuration.}
\label{tab:rtxpro_time}
\resizebox{0.9\linewidth}{!}{%
\begin{tabular}{|c|l!{\vrule width 1.6pt}c|c!{\vrule width 1.2pt}c|c!{\vrule width 1.6pt}c|c!{\vrule width 1.2pt}c|c!{\vrule width 1.6pt}c|c!{\vrule width 1.2pt}c|c!{\vrule width 1.6pt}c|c!{\vrule width 1.2pt}c|c!{\vrule width 1.6pt}}
\hline
\multicolumn{2}{|c!{\vrule width 1.6pt}}{Radius distribution (r)} & \multicolumn{4}{c!{\vrule width 1.6pt}}{r=1} & \multicolumn{4}{c!{\vrule width 1.6pt}}{r=160} & \multicolumn{4}{c!{\vrule width 1.6pt}}{r=(U, {[}1, 160{]})} & \multicolumn{4}{c!{\vrule width 1.6pt}}{r=(LN, {[}1, 330{]})} \\
\hline
\multicolumn{2}{|c!{\vrule width 1.6pt}}{Boundary conditions (BC)} & \multicolumn{2}{c!{\vrule width 1.2pt}}{Wall} & \multicolumn{2}{c!{\vrule width 1.6pt}}{Periodic} & \multicolumn{2}{c!{\vrule width 1.2pt}}{Wall} & \multicolumn{2}{c!{\vrule width 1.6pt}}{Periodic} & \multicolumn{2}{c!{\vrule width 1.2pt}}{Wall} & \multicolumn{2}{c!{\vrule width 1.6pt}}{Periodic} & \multicolumn{2}{c!{\vrule width 1.2pt}}{Wall} & \multicolumn{2}{c!{\vrule width 1.6pt}}{Periodic} \\
\hline
\multicolumn{2}{|c!{\vrule width 1.6pt}}{Number of particles (n)} & \multicolumn{1}{c|}{50k} & \multicolumn{1}{c!{\vrule width 1.2pt}}{1M} & \multicolumn{1}{c|}{50k} & \multicolumn{1}{c!{\vrule width 1.6pt}}{1M} & \multicolumn{1}{c|}{50k} & \multicolumn{1}{c!{\vrule width 1.2pt}}{1M} & \multicolumn{1}{c|}{50k} & \multicolumn{1}{c!{\vrule width 1.6pt}}{1M} & \multicolumn{1}{c|}{50k} & \multicolumn{1}{c!{\vrule width 1.2pt}}{1M} & \multicolumn{1}{c|}{50k} & \multicolumn{1}{c!{\vrule width 1.6pt}}{1M} & \multicolumn{1}{c|}{50k} & \multicolumn{1}{c!{\vrule width 1.2pt}}{1M} & \multicolumn{1}{c|}{50k} & \multicolumn{1}{c!{\vrule width 1.6pt}}{1M} \\
\hline
\thickhline
\multicolumn{1}{|c|}{\multirow{5}{*}{Lattice}} & CPU-CELL@64c & 1.80 & 9.64 & 1.80 & 10.4 & 17.9 & 5482 & \cellcolor{FastTeal!22}28.0 & 9864 & 11.7 & 3958 & 21.2 & 7891 & 56.4 & 22311 & 141 & 60612 \\
\cline{2-18}
\multicolumn{1}{|c|}{} & GPU-CELL & 0.415 & 2.97 & 0.399 & 2.61 & 43.6 & 3185 & 38.4 & \cellcolor{FastTeal!22}5379 & 27.1 & 2163 & 26.3 & 3957 & 130 & 7767 & 151 & 25864 \\
\cline{2-18}
\multicolumn{1}{|c|}{} & RT-REF & 0.165 & 0.599 & 0.199 & 0.844 & \cellcolor{FastTeal!22}17.6 & \cellcolor{FastTeal!22}2235 & 31.8 & 5520 & 6.99 & \cellcolor{FastTeal!22}835 & 10.9 & \cellcolor{FastTeal!22}1801 & 10.6 & - & 14.8 & - \\
\cline{2-18}
\multicolumn{1}{|c|}{} & ORCS-forces & 0.166 & 0.612 & 0.181 & 0.729 & 19.5 & 4836 & 35.8 & 7844 & \cellcolor{FastTeal!22}4.99 & 1323 & \cellcolor{FastTeal!22}7.99 & 2064 & \cellcolor{FastTeal!22}2.33 & \cellcolor{FastTeal!22}548 & \cellcolor{FastTeal!22}4.26 & \cellcolor{FastTeal!22}976 \\
\cline{2-18}
\multicolumn{1}{|c|}{} & ORCS-perse & \cellcolor{FastTeal!22}0.127 & \cellcolor{FastTeal!22}0.506 & \cellcolor{FastTeal!22}0.147 & \cellcolor{FastTeal!22}0.632 & 19.6 & 5201 & 36.4 & 8674 & - & - & - & - & - & - & - & - \\
\thickhline
\multicolumn{1}{|c|}{\multirow{5}{*}{Disordered}} & CPU-CELL@64c & 1.97 & 13.7 & 2.01 & 14.2 & 21.6 & 8229 & \cellcolor{FastTeal!22}31.5 & 12156 & 12.0 & 4221 & 21.2 & 7935 & 56.0 & 22204 & 268 & 106472 \\
\cline{2-18}
\multicolumn{1}{|c|}{} & GPU-CELL & 0.433 & 3.49 & 0.419 & 3.08 & 48.6 & 4569 & 42.5 & \cellcolor{FastTeal!22}7174 & 28.4 & 2697 & 27.7 & 4664 & 131 & 8073 & 156 & 26890 \\
\cline{2-18}
\multicolumn{1}{|c|}{} & RT-REF & 0.163 & 0.658 & 0.192 & 1.03 & \cellcolor{FastTeal!22}18.6 & \cellcolor{FastTeal!22}2861 & 39.7 & 7525 & 7.08 & \cellcolor{FastTeal!22}972 & 12.4 & \cellcolor{FastTeal!22}2250 & 8.73 & - & 14.9 & - \\
\cline{2-18}
\multicolumn{1}{|c|}{} & ORCS-forces & 0.153 & 0.660 & 0.173 & 0.894 & 22.3 & 5398 & 45.7 & 10989 & \cellcolor{FastTeal!22}5.41 & 1464 & \cellcolor{FastTeal!22}9.95 & 2671 & \cellcolor{FastTeal!22}2.59 & \cellcolor{FastTeal!22}674 & \cellcolor{FastTeal!22}5.60 & \cellcolor{FastTeal!22}1484 \\
\cline{2-18}
\multicolumn{1}{|c|}{} & ORCS-perse & \cellcolor{FastTeal!22}0.117 & \cellcolor{FastTeal!22}0.552 & \cellcolor{FastTeal!22}0.137 & \cellcolor{FastTeal!22}0.792 & 22.5 & 6017 & 47.0 & 12573 & - & - & - & - & - & - & - & - \\
\thickhline
\multicolumn{1}{|c|}{\multirow{5}{*}{Cluster}} & CPU-CELL@64c & 1.52 & 15.4 & 1.56 & 18.1 & \cellcolor{FastTeal!22}153 & 13820 & \cellcolor{FastTeal!22}147 & 17827 & \cellcolor{FastTeal!22}111 & 10969 & \cellcolor{FastTeal!22}194 & 21459 & 66.8 & 21617 & 188 & 95351 \\
\cline{2-18}
\multicolumn{1}{|c|}{} & GPU-CELL & 0.638 & 6.82 & 0.617 & 7.62 & 333 & \cellcolor{FastTeal!22}7119 & 201 & \cellcolor{FastTeal!22}9392 & 278 & \cellcolor{FastTeal!22}6114 & 302 & \cellcolor{FastTeal!22}10278 & 229 & 12241 & 318 & 30759 \\
\cline{2-18}
\multicolumn{1}{|c|}{} & RT-REF & 0.200 & \cellcolor{FastTeal!22}4.77 & 0.227 & \cellcolor{FastTeal!22}7.43 & 251 & - & 354 & - & 182 & - & 342 & - & 145 & - & 295 & - \\
\cline{2-18}
\multicolumn{1}{|c|}{} & ORCS-forces & 0.176 & 6.32 & 0.186 & 7.82 & 223 & 19150 & 281 & 57670 & 119 & 6796 & 202 & 13780 & \cellcolor{FastTeal!22}40.8 & \cellcolor{FastTeal!22}2586 & \cellcolor{FastTeal!22}75.0 & \cellcolor{FastTeal!22}5617 \\
\cline{2-18}
\multicolumn{1}{|c|}{} & ORCS-perse & \cellcolor{FastTeal!22}0.146 & 7.44 & \cellcolor{FastTeal!22}0.158 & 8.99 & 232 & 23445 & 301 & 78784 & - & - & - & - & - & - & - & - \\
\thickhline
\end{tabular}%
}
\end{table*}

\begin{figure*}[ht!]
    \centering
    \includegraphics[width=0.24\linewidth]{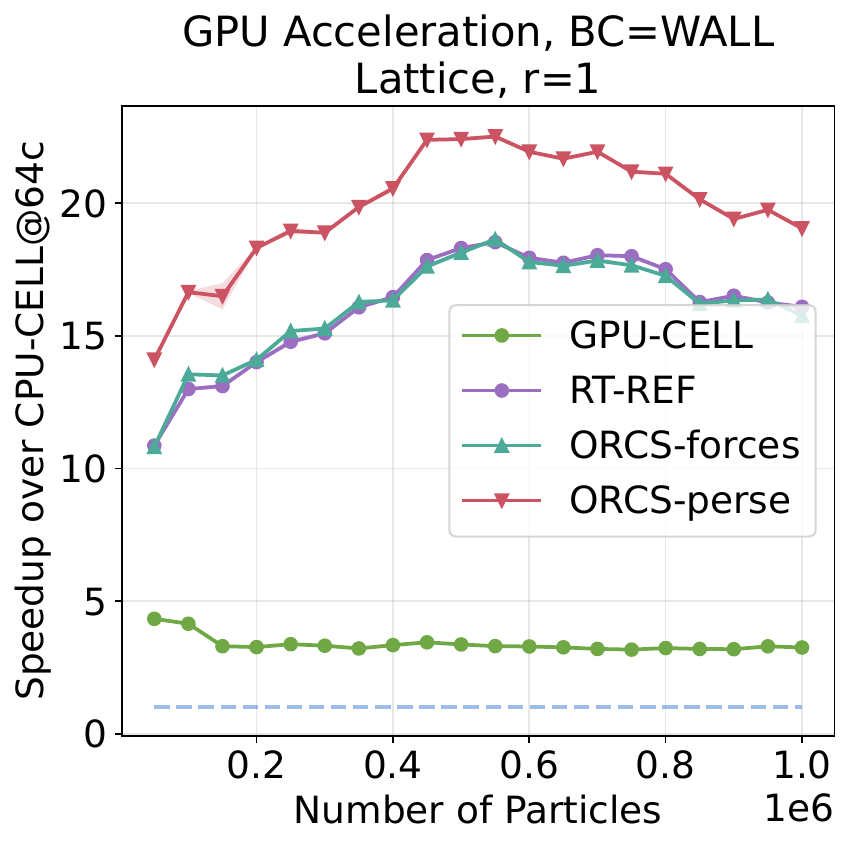}
    \includegraphics[width=0.24\linewidth]{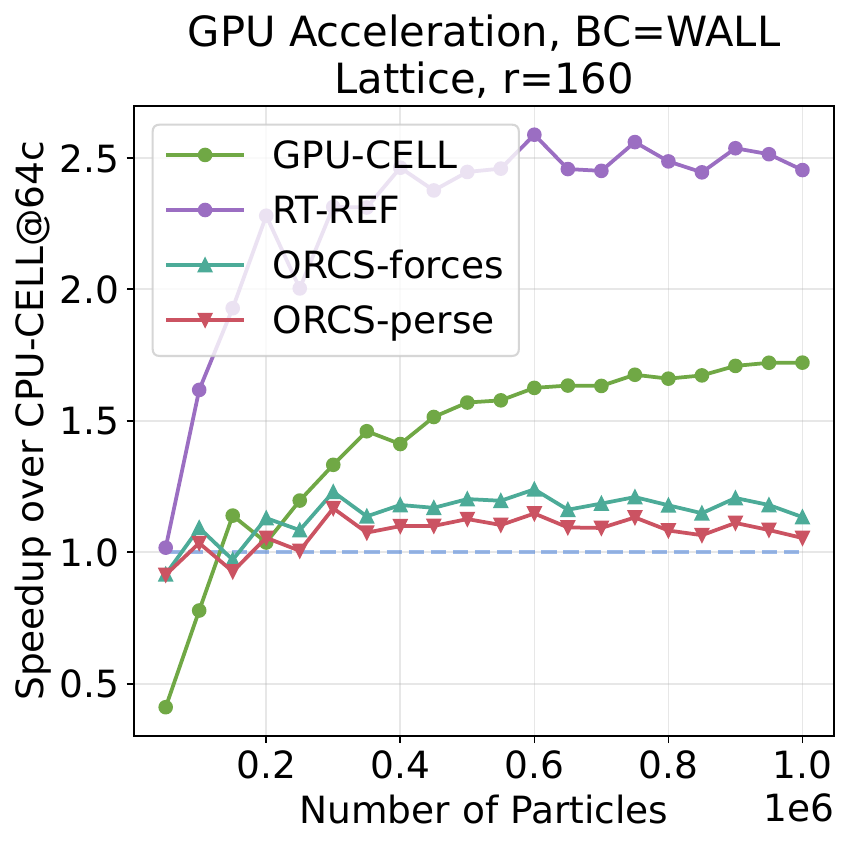}
    \includegraphics[width=0.24\linewidth]{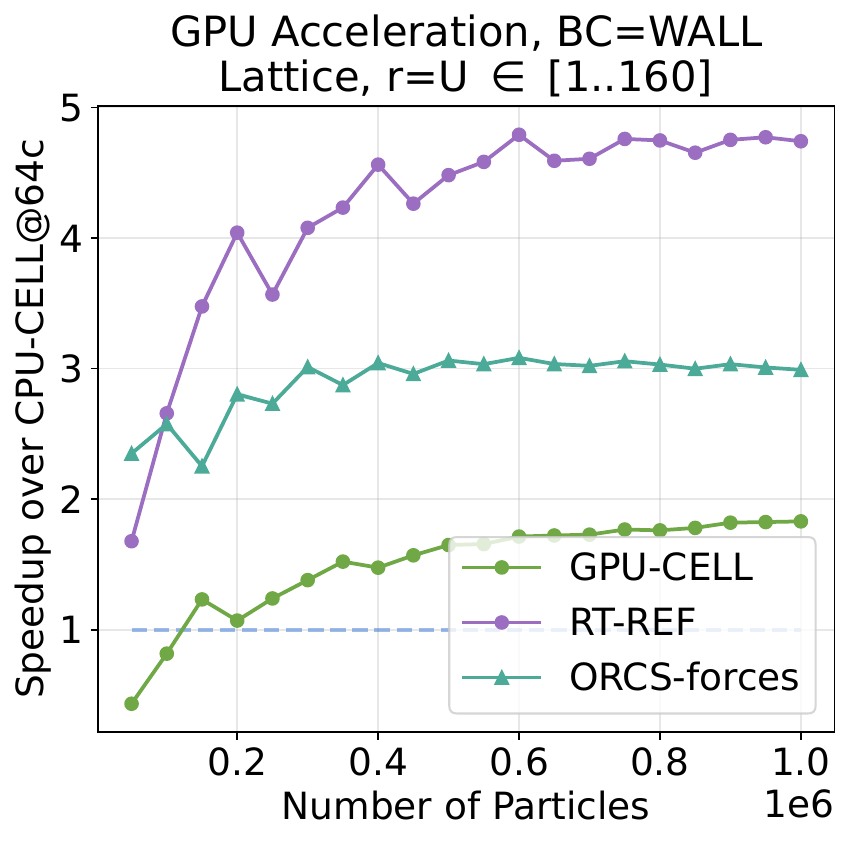}
    \includegraphics[width=0.24\linewidth]{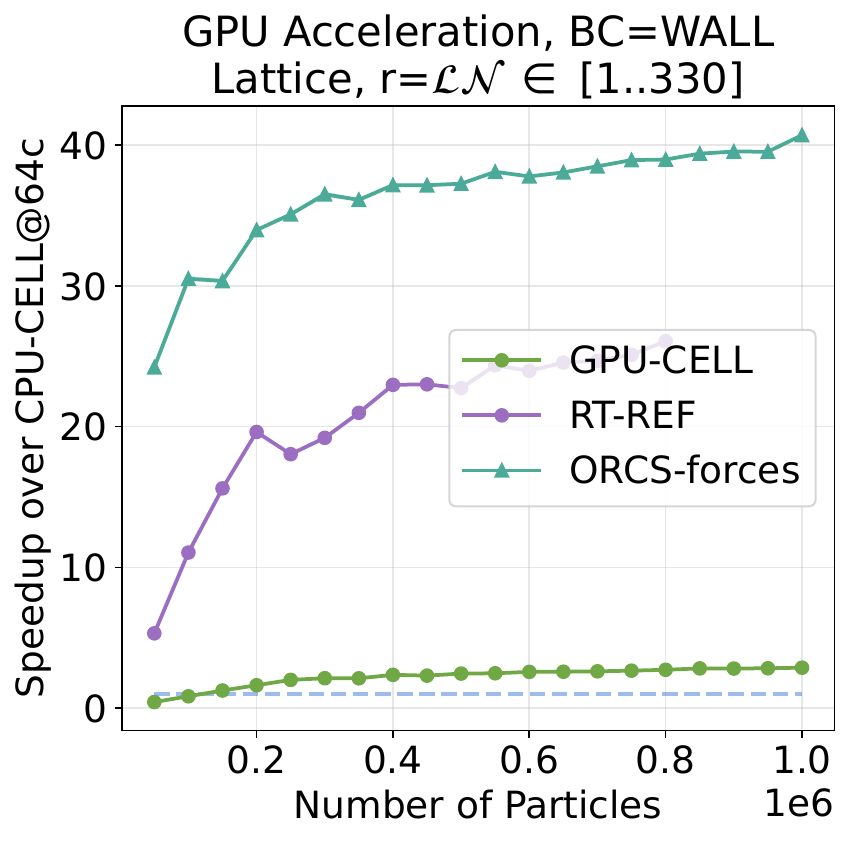}\\
    
    \includegraphics[width=0.24\linewidth]{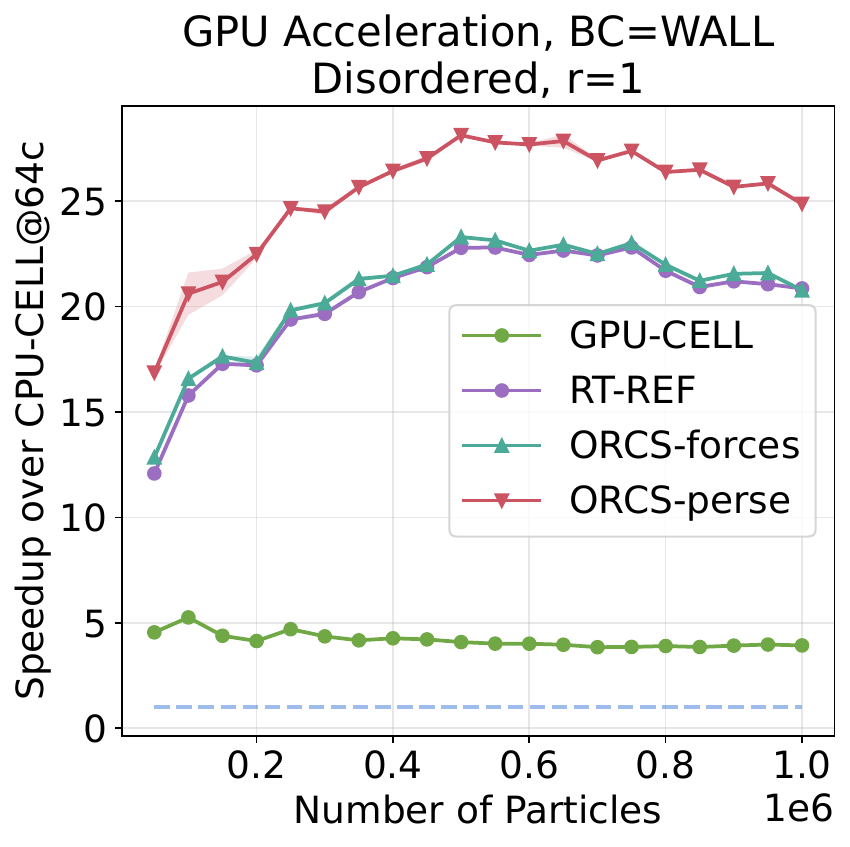}
    \includegraphics[width=0.24\linewidth]{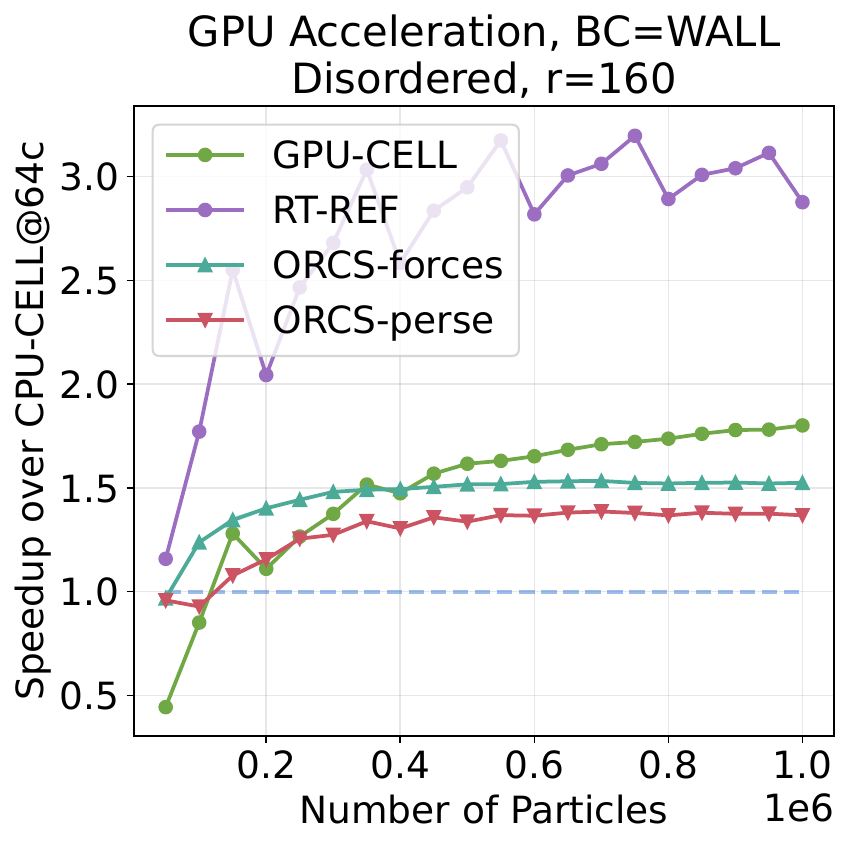}
    \includegraphics[width=0.24\linewidth]{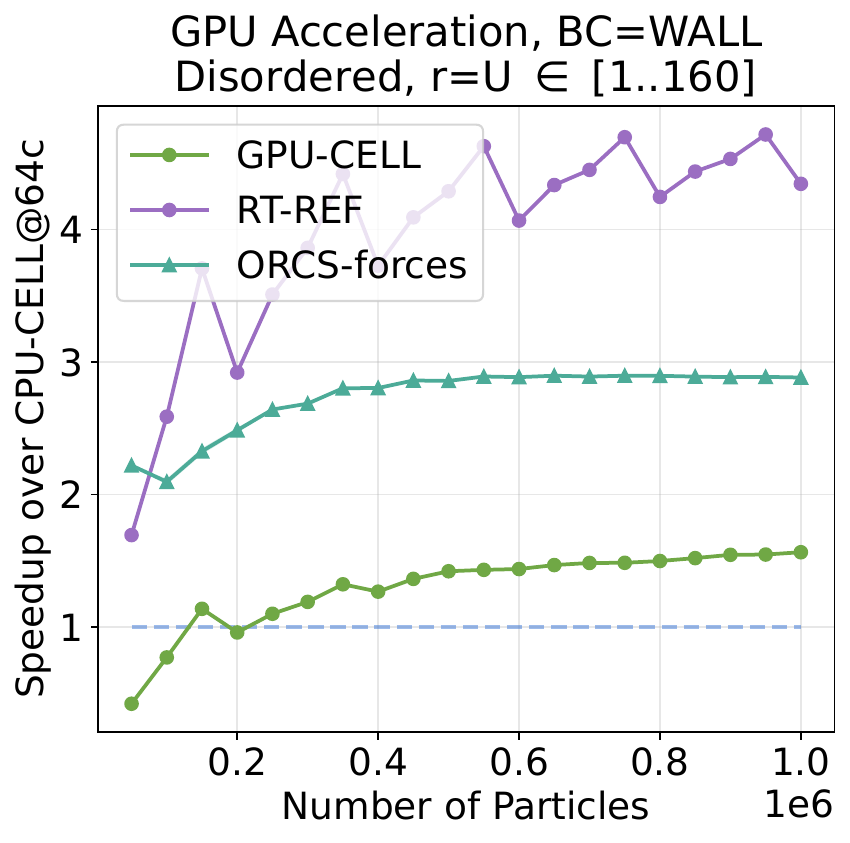}
    \includegraphics[width=0.24\linewidth]{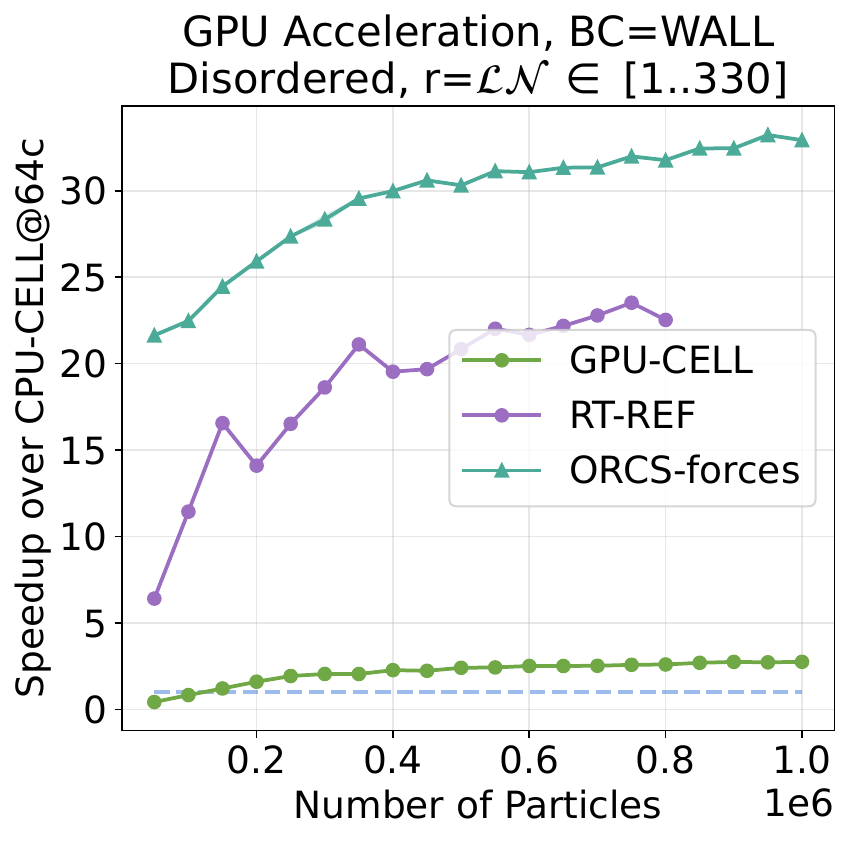}\\
    
    \includegraphics[width=0.24\linewidth]{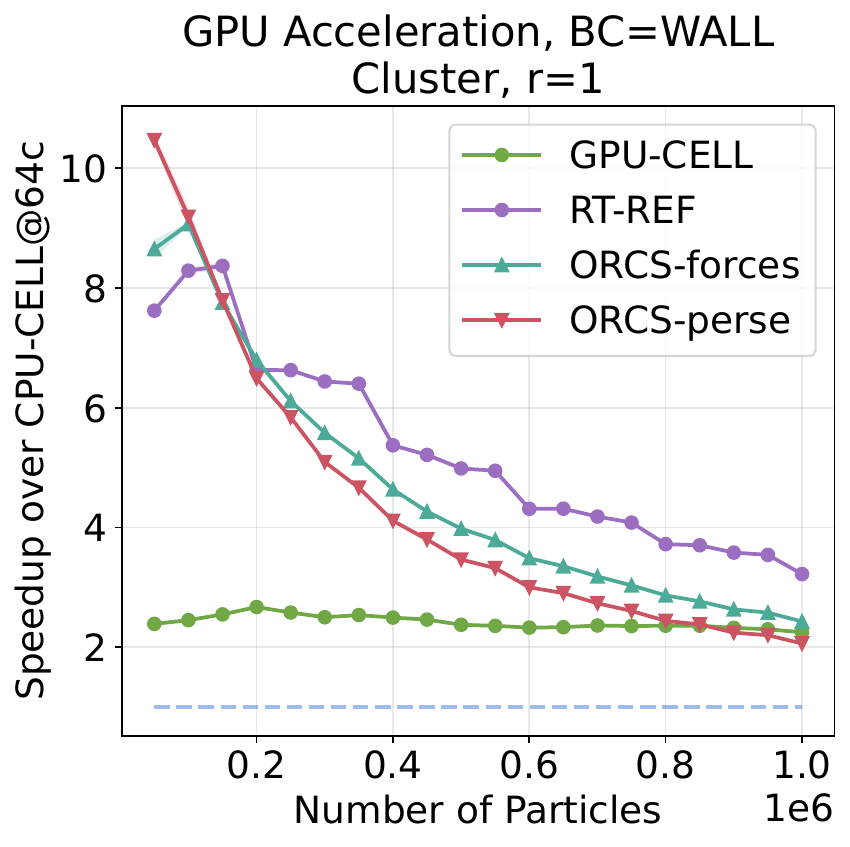}
    \includegraphics[width=0.24\linewidth]{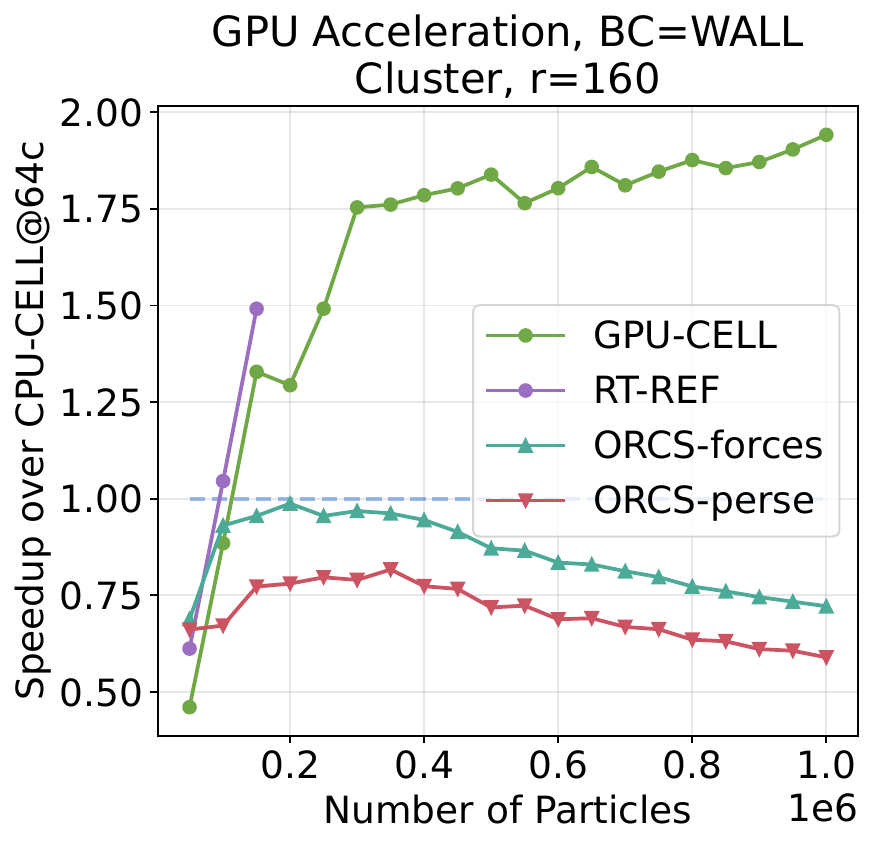}
    \includegraphics[width=0.24\linewidth]{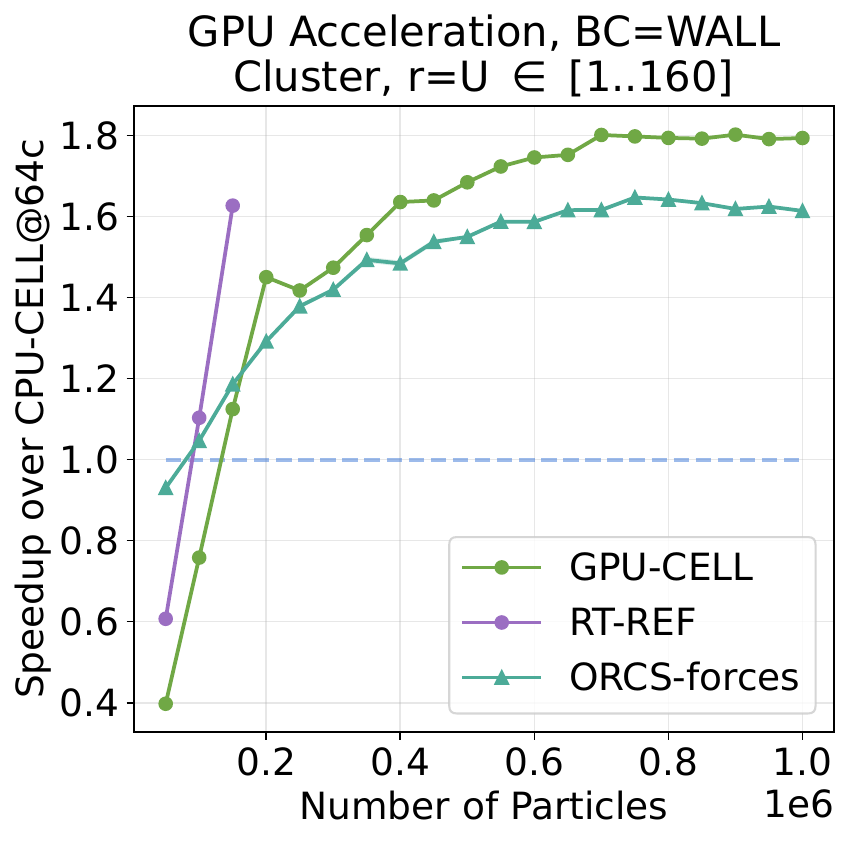}
    \includegraphics[width=0.24\linewidth]{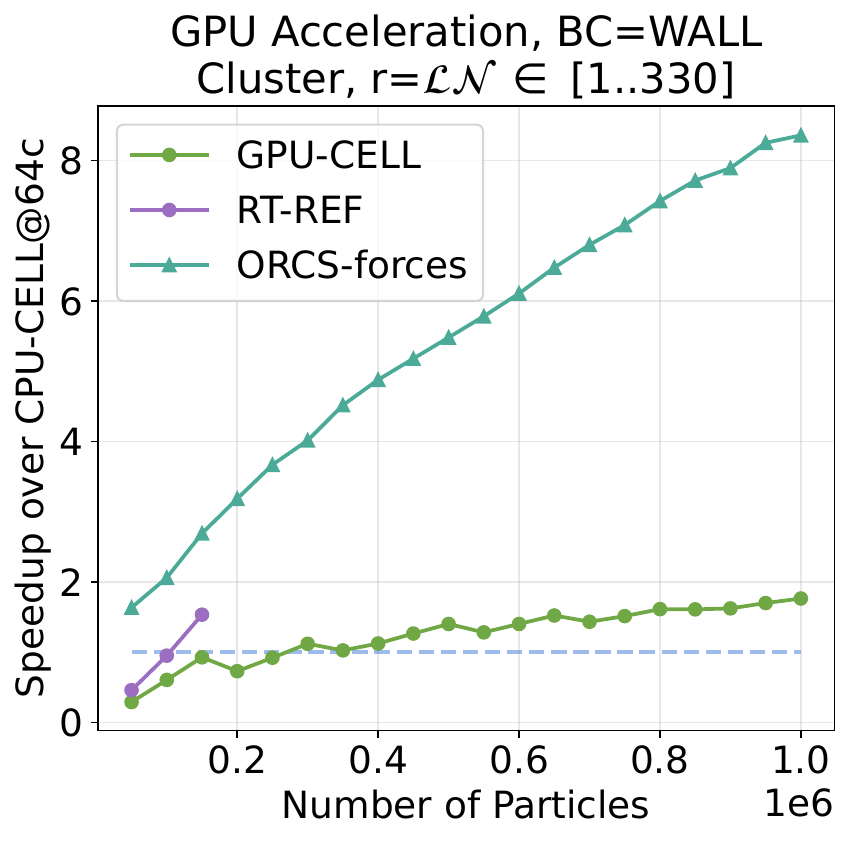}
    \caption{GPU acceleration with respect to \texttt{CPU-CELL@64c} using WALL boundary conditions. Three different particle distributions are employed; Lattice, Disordered and Cluster, under four different radius distributions; fixed $r=1$, fixed $r=160$, random uniform (U) $\in [1,160]$ and LogNormal ($\mathcal{LN}$) $\in [1, 330]$.}
    \label{plot:speedup-WALL}
\end{figure*}

Figure \ref{plot:speedup-WALL} presents the GPU acceleration of the different GPU approaches with respect to the parallel \texttt{CPU-CELL@64c}, for an increasing number of particles using Wall BC. Here, speedup is the quotient of:
\begin{equation}
    \text{Speedup} = \frac{\langle T_{\text{CPU-CELL@64c}}\rangle }{\langle T_{\text{GPU}}\rangle}    
\end{equation}
where $\langle T \rangle$ is the average time per simulation step.
One pattern to notice is that \texttt{ORCS-pers\'e} and \texttt{ORCS-forces} are the fastest approaches at low (first column) and logNormal (last column) radius distributions, respectively. In particular, in the first column \texttt{ORCS-pers\'e} can run up to $\sim 24\times$ and $\sim 29\times$ faster than the CPU approach in the Lattice and Disordered cases, respectively, while the base \texttt{RT-REF} and \texttt{ORCS-forces} reach up to $\sim 18\times$ and $\sim 24\times$ for the same cases. For reference, \texttt{GPU-CELL} only reaches up to $\sim 5\times$ of speedup on those cases. This makes the proposed variant roughly $\sim1.3\times$ faster than the \texttt{RT-REF}. In the last column, \texttt{ORCS-forces} runs up to $\sim40\times, \sim35\times$ and $\sim8\times$ faster than \texttt{CPU-CELL@64c}, followed by \texttt{RT-REF} which in its best case achieves up to $25\times$ of speedup. This puts \texttt{ORCS-forces} up to $\sim1.6\times$ faster than \texttt{RT-REF}. Next comes \texttt{GPU-CELL} with up to $3\times$ of speedup in the best case. When Lattice and Disordered distributions are combined with medium-large radius such as $r=160$ or $r \in [1,160]$, \texttt{RT-REF} is the fastest approach. However, it runs out of memory when simulating the Cluster particle distribution with a log normal radius distribution, because its neighbor list must cover the worst case. In the case of \texttt{GPU-CELL}, it is highly competitive when the grid cells are balanced in size, such as when $r=160$, and is the less affected approach when the number of radius interactions is extreme as in the Cluster $r=160$ case. In fact, this case has so much interactions that the two proposed RT Core solutions perform slower than the CPU approach, while the \texttt{RT-REF} runs out of memory early in $n$. 

\begin{figure*}[ht!]
    \centering
    \includegraphics[width=0.24\linewidth]{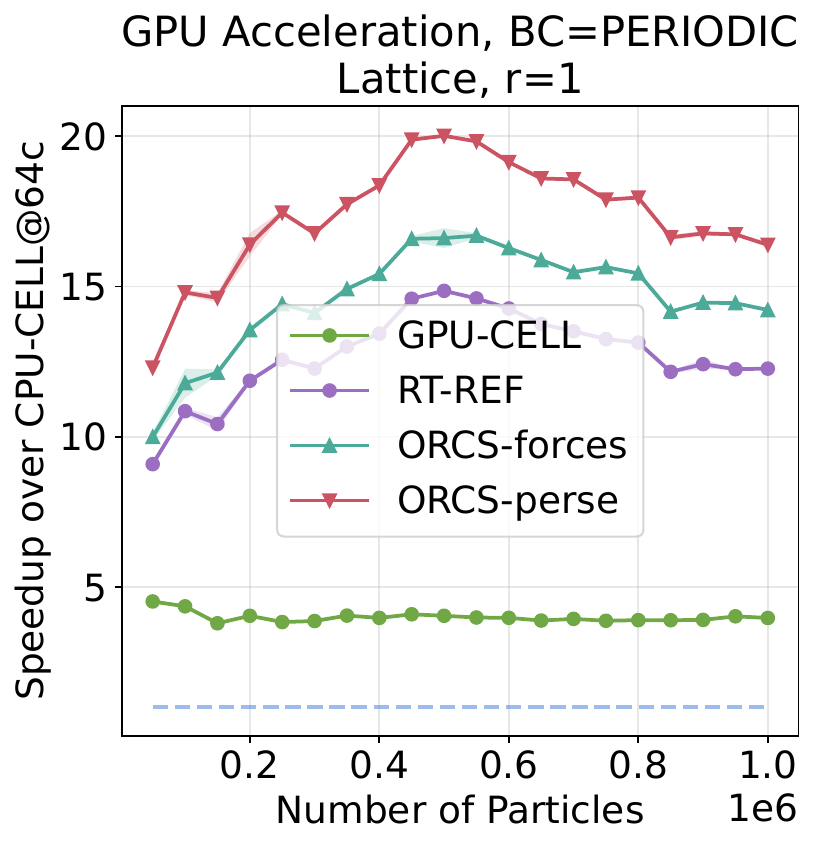}
    \includegraphics[width=0.24\linewidth]{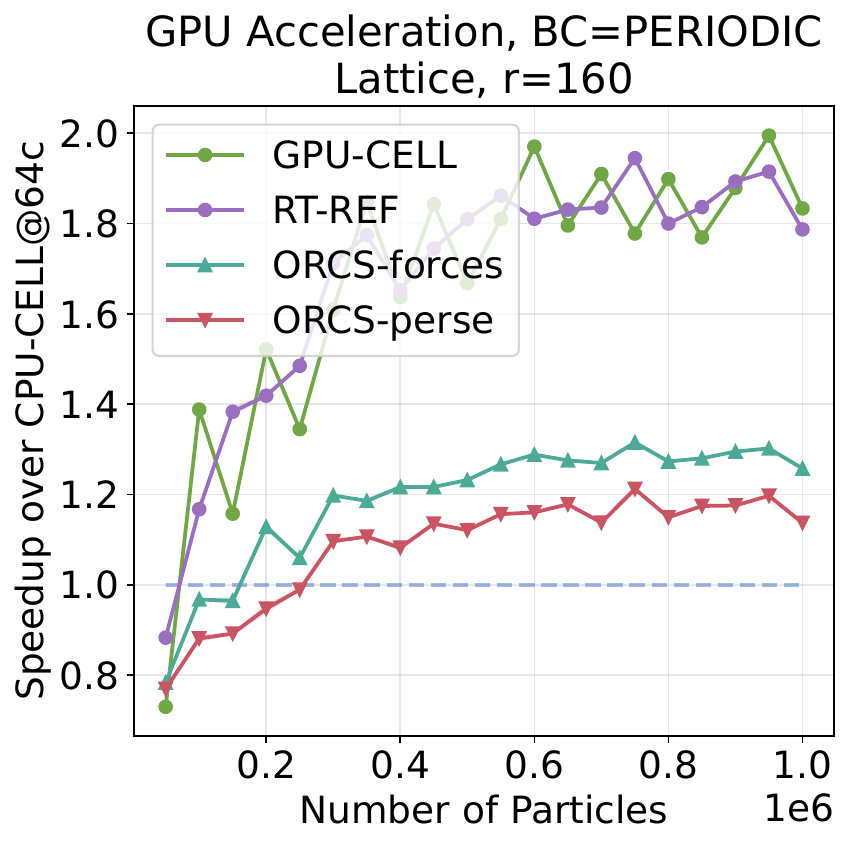}
    \includegraphics[width=0.24\linewidth]{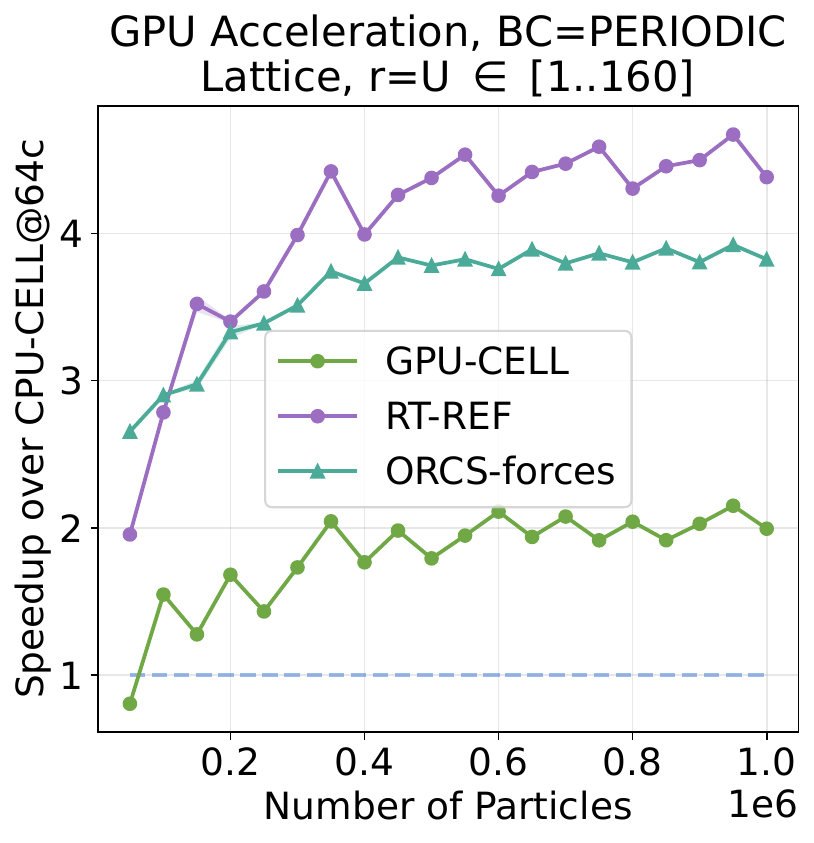}
    \includegraphics[width=0.24\linewidth]{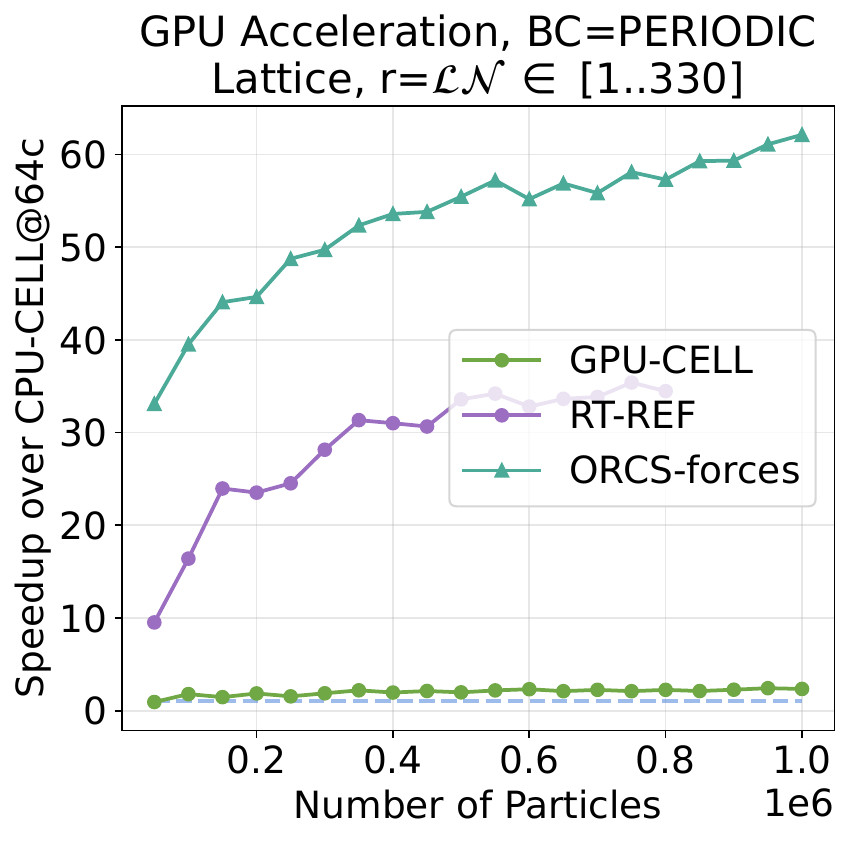}\\
    
    \includegraphics[width=0.24\linewidth]{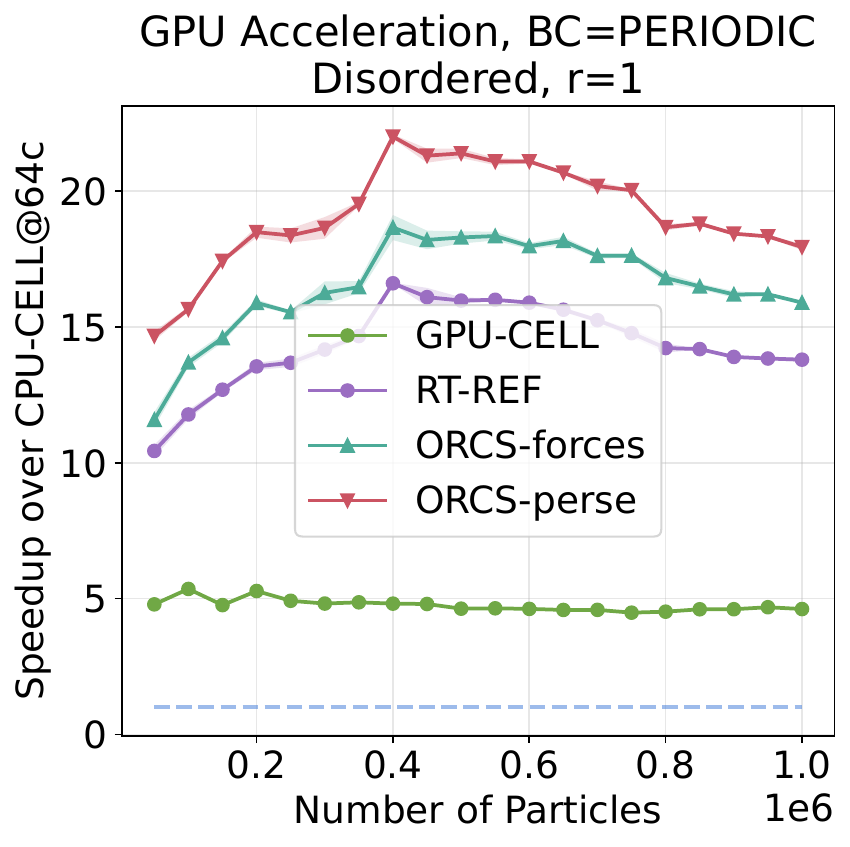}
    \includegraphics[width=0.24\linewidth]{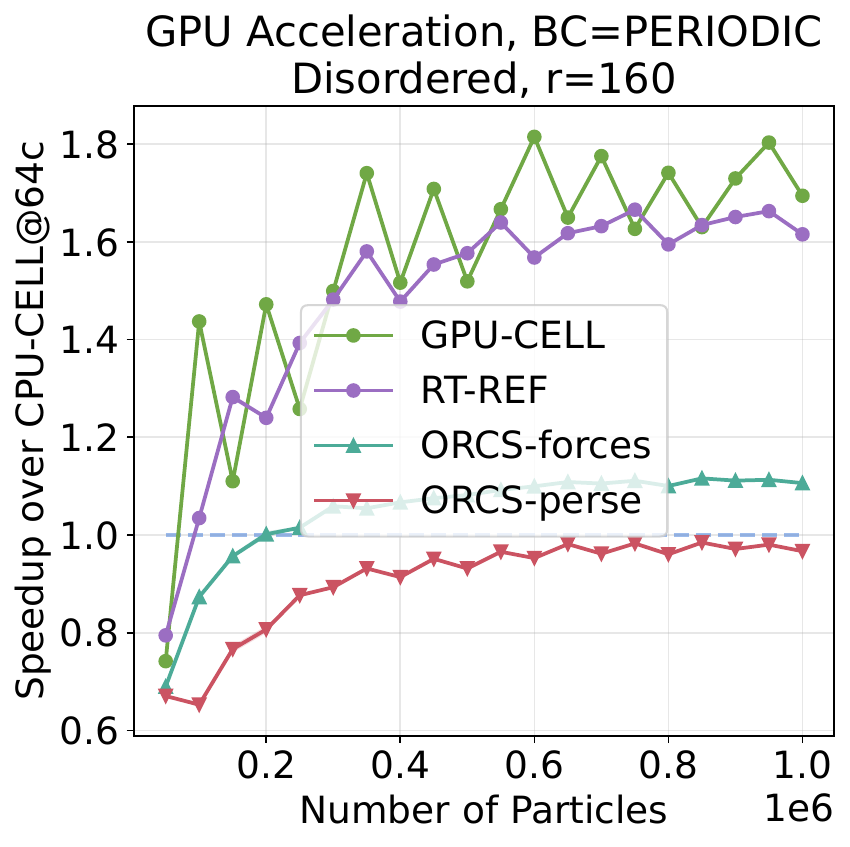}
    \includegraphics[width=0.24\linewidth]{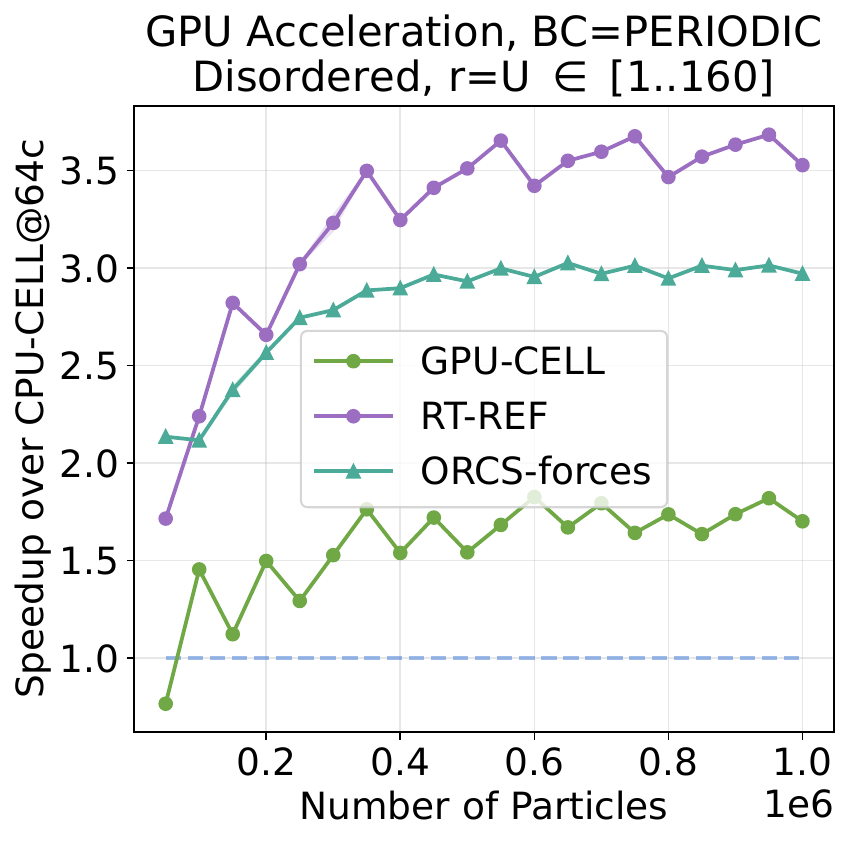}
    \includegraphics[width=0.24\linewidth]{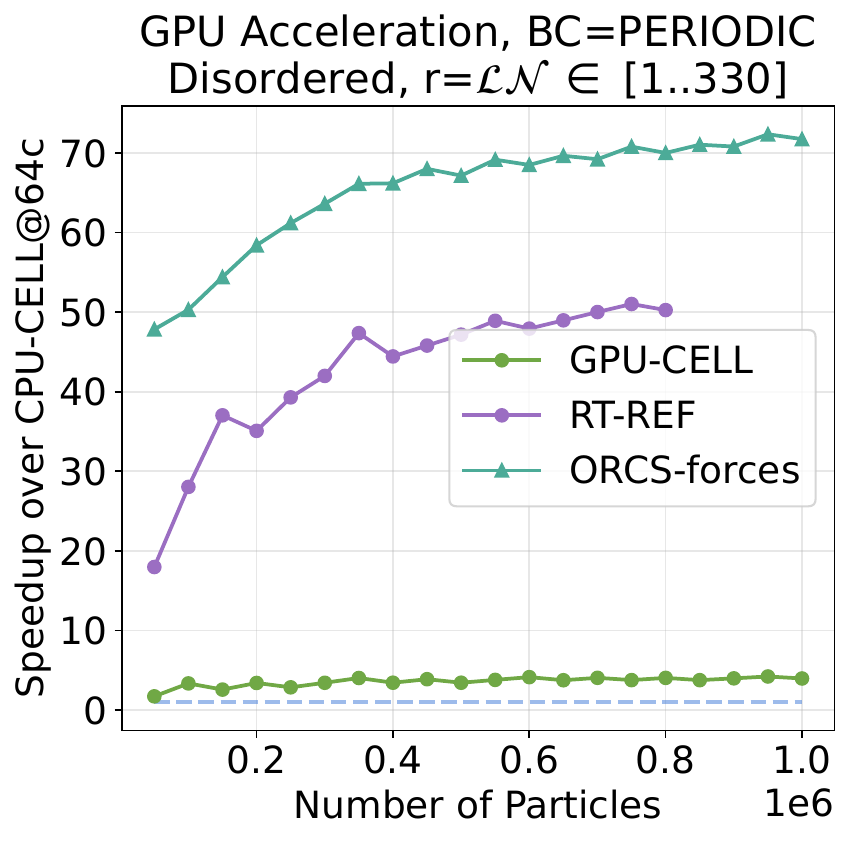}\\
    
    \includegraphics[width=0.24\linewidth]{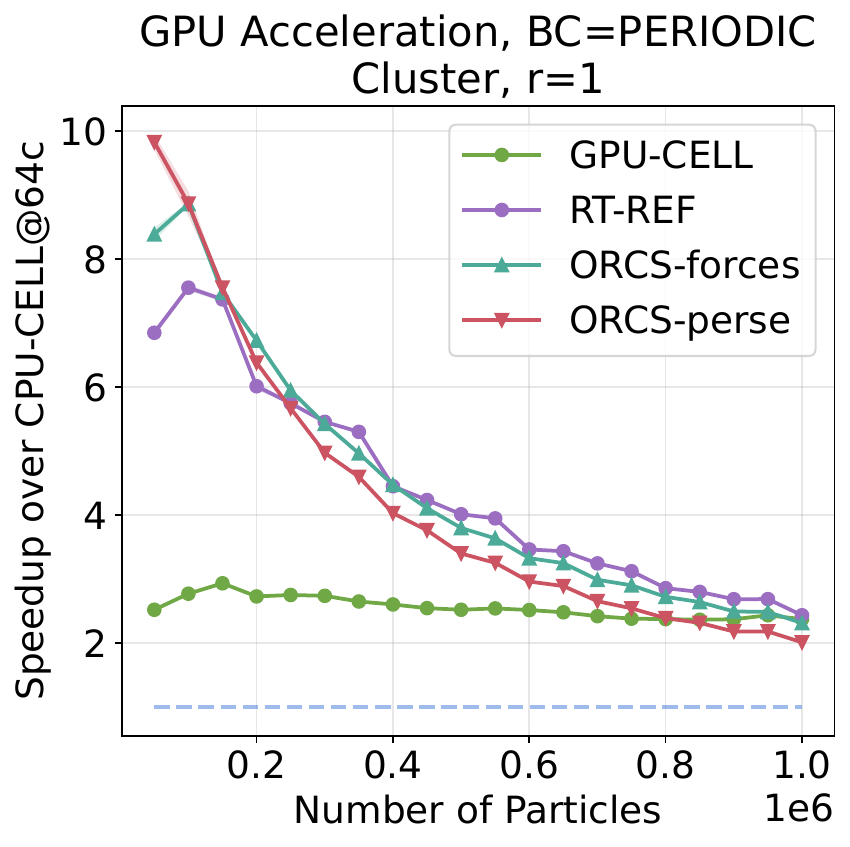}
    \includegraphics[width=0.24\linewidth]{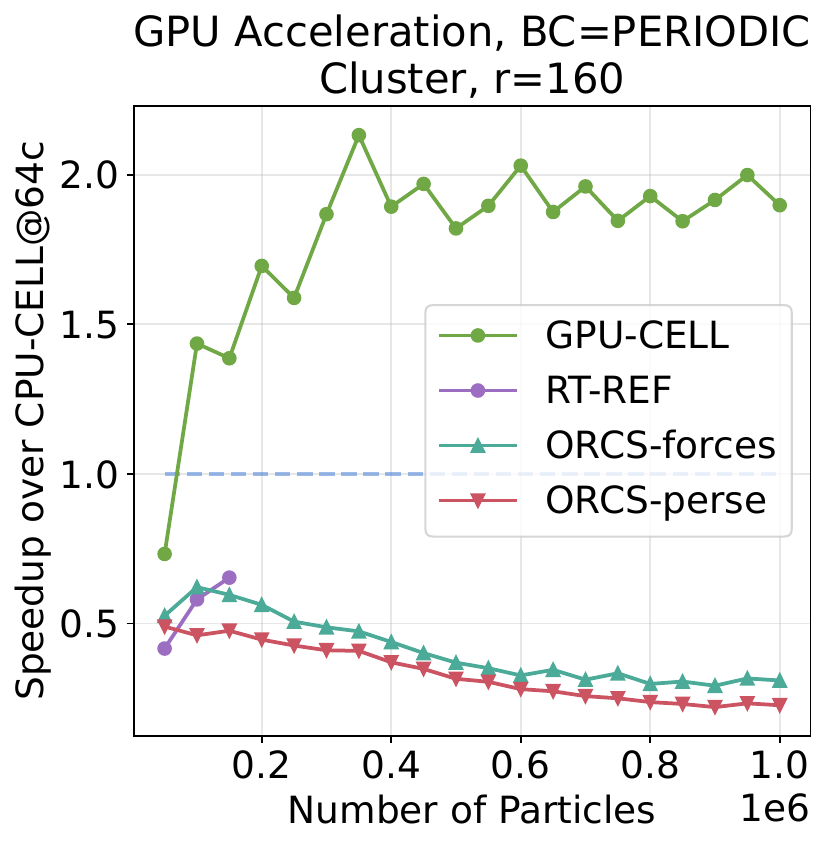}
    \includegraphics[width=0.24\linewidth]{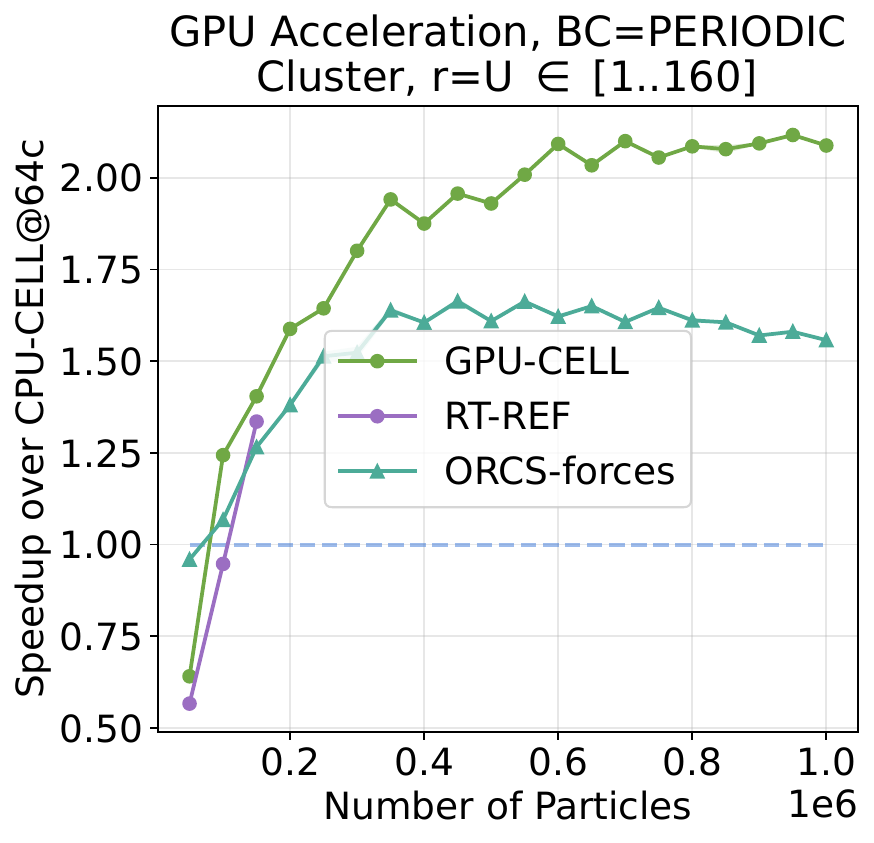}
    \includegraphics[width=0.24\linewidth]{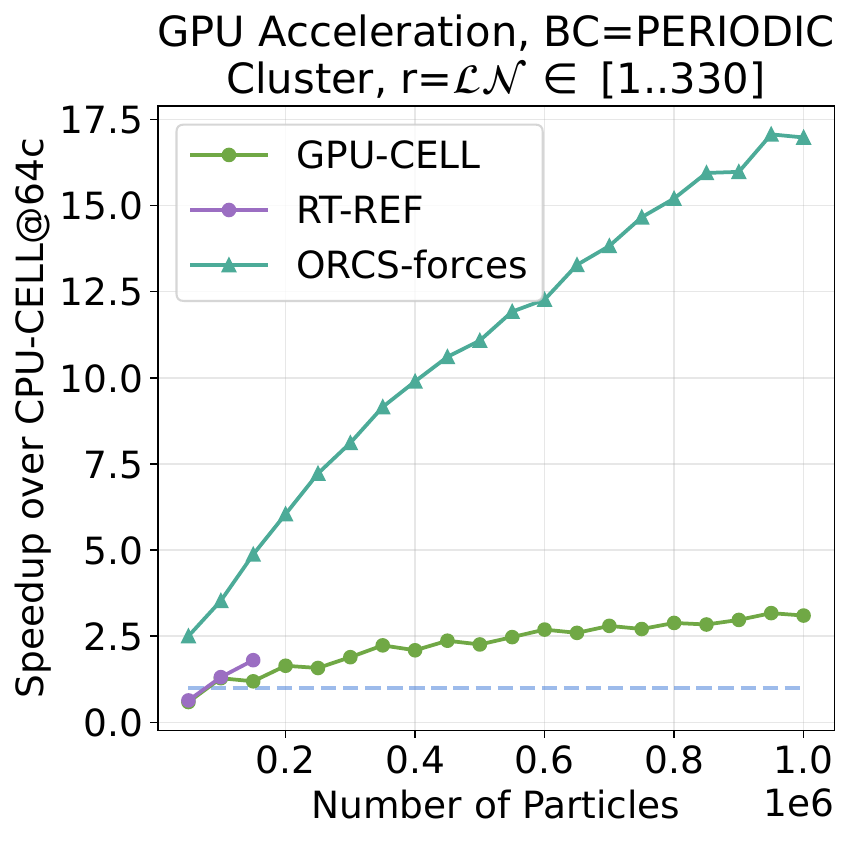}
    \caption{GPU acceleration with respect to \texttt{CPU-CELL@64c} using PERIODIC boundary conditions. Three different particle distributions are employed; Lattice, Disordered and Cluster, under four different radius distributions; fixed $r=1$, fixed $r=160$, random uniform (U) $\in [1,160]$ and LogNormal ($\mathcal{LN}$) $\in [1, 330]$.}
    \label{plot:speedup-PERIODIC}
\end{figure*}

Figure \ref{plot:speedup-PERIODIC} presents the speedup results for the case of Periodic boundary conditions. Although the pattern is highly similar to wall BC, in the first two columns the plots reach lower speedup values, the third column stays with little changes, and the last column (log normal radius distribution) shows much higher speedups in favor of the RT Core approaches, specially for the proposed \texttt{ORCS-forces} which manages to be near $\sim 2\times$ faster than \texttt{RT-REF}.

\subsection{Energy Efficiency and Scaling}
\begin{figure*}[ht!]
    \centering
    \includegraphics[width=0.32\linewidth]{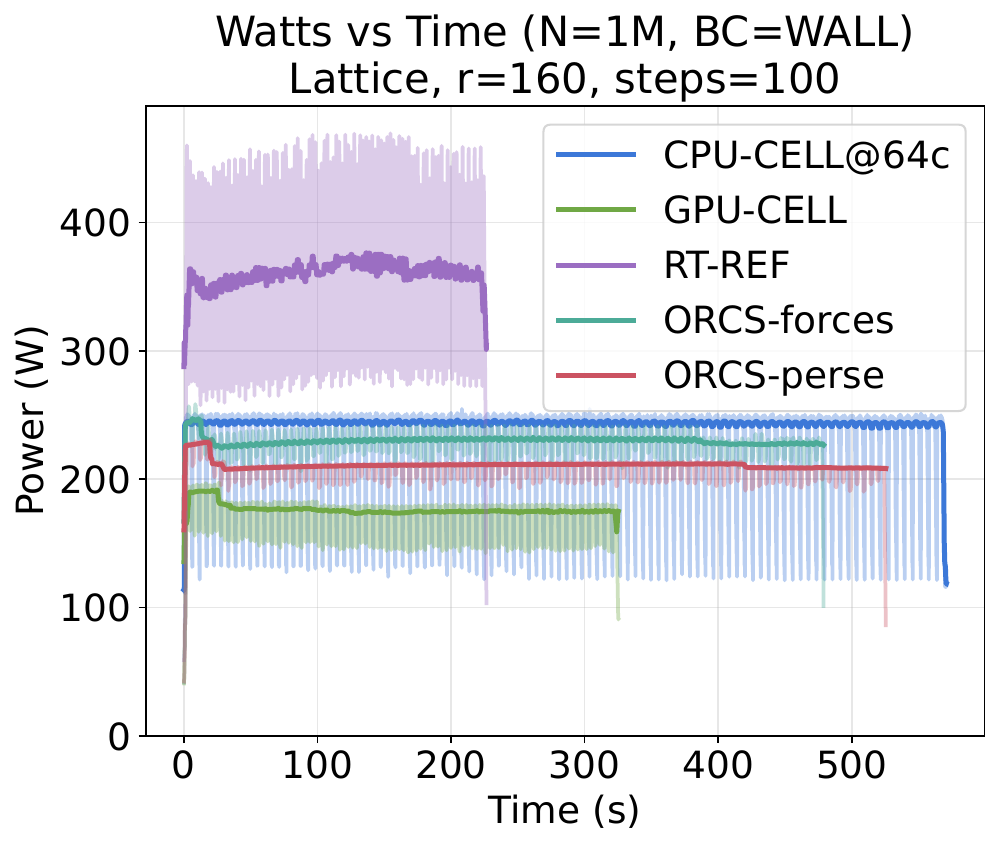}
    \includegraphics[width=0.32\linewidth]{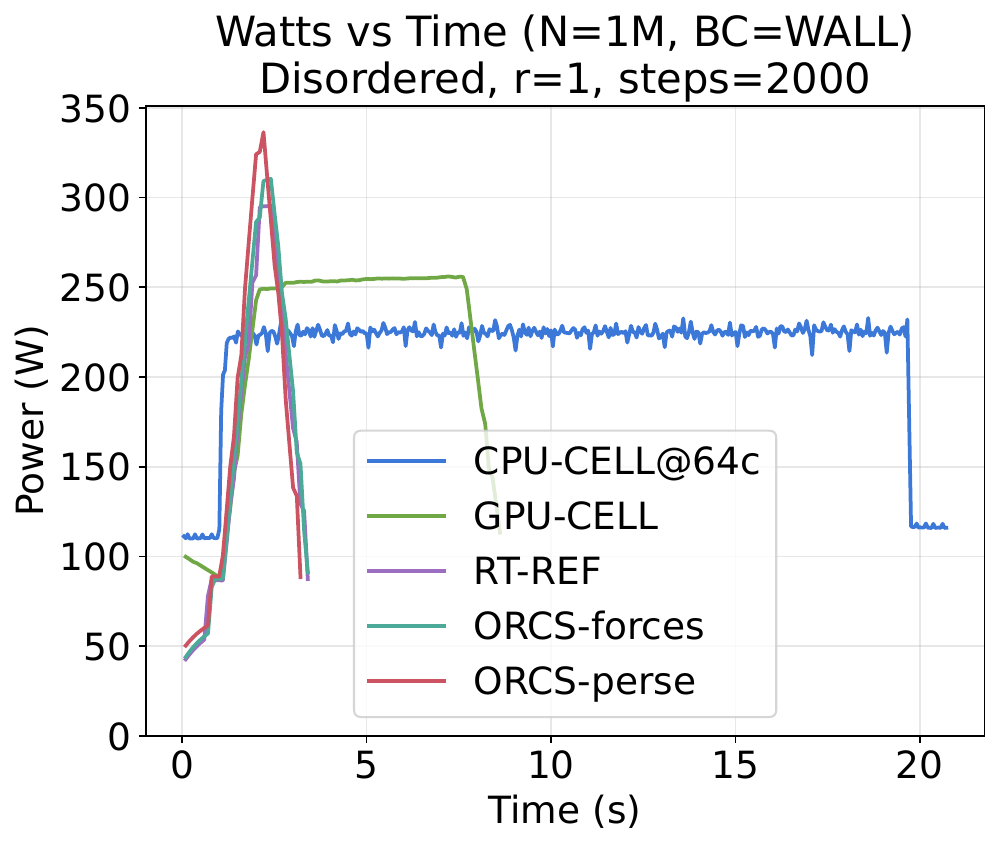}
    \includegraphics[width=0.32\linewidth]{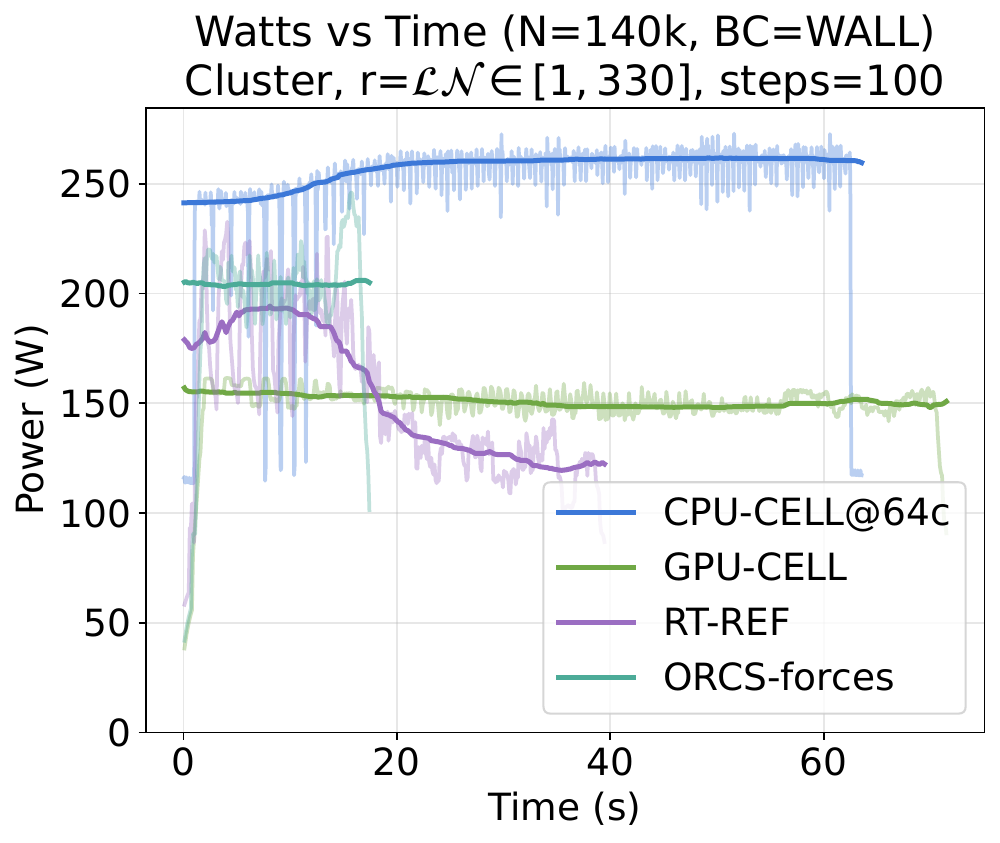}\\
    
    \includegraphics[width=0.32\linewidth]{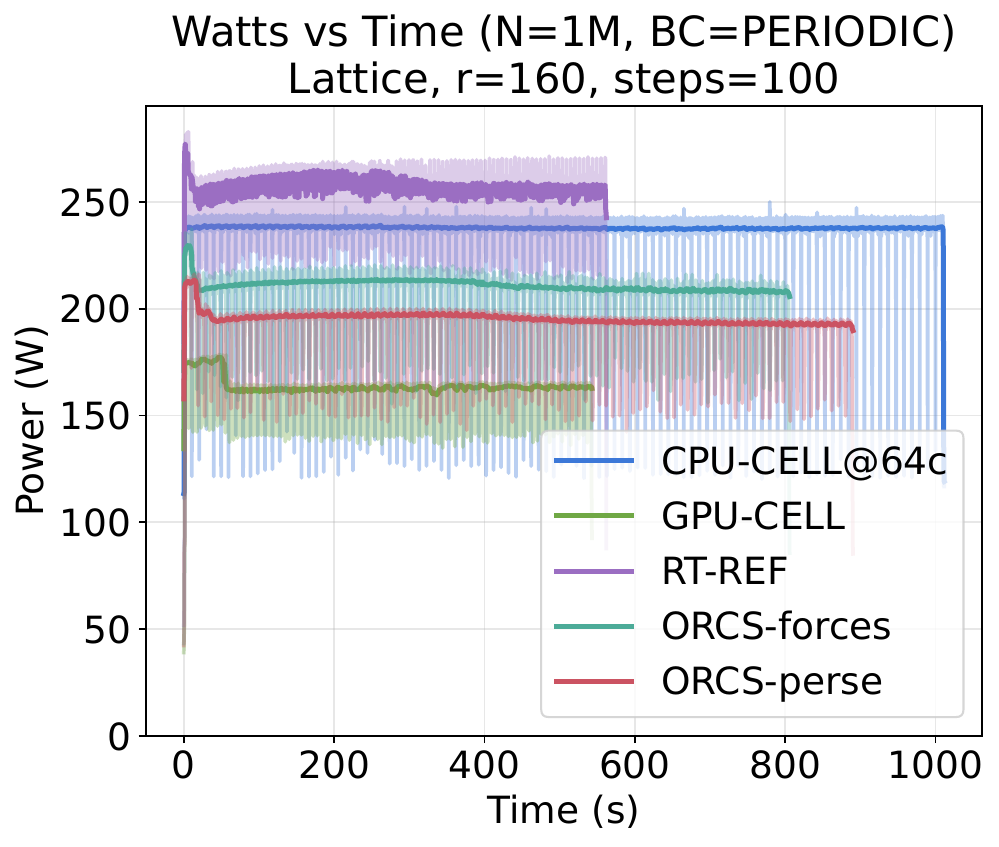}
    \includegraphics[width=0.32\linewidth]{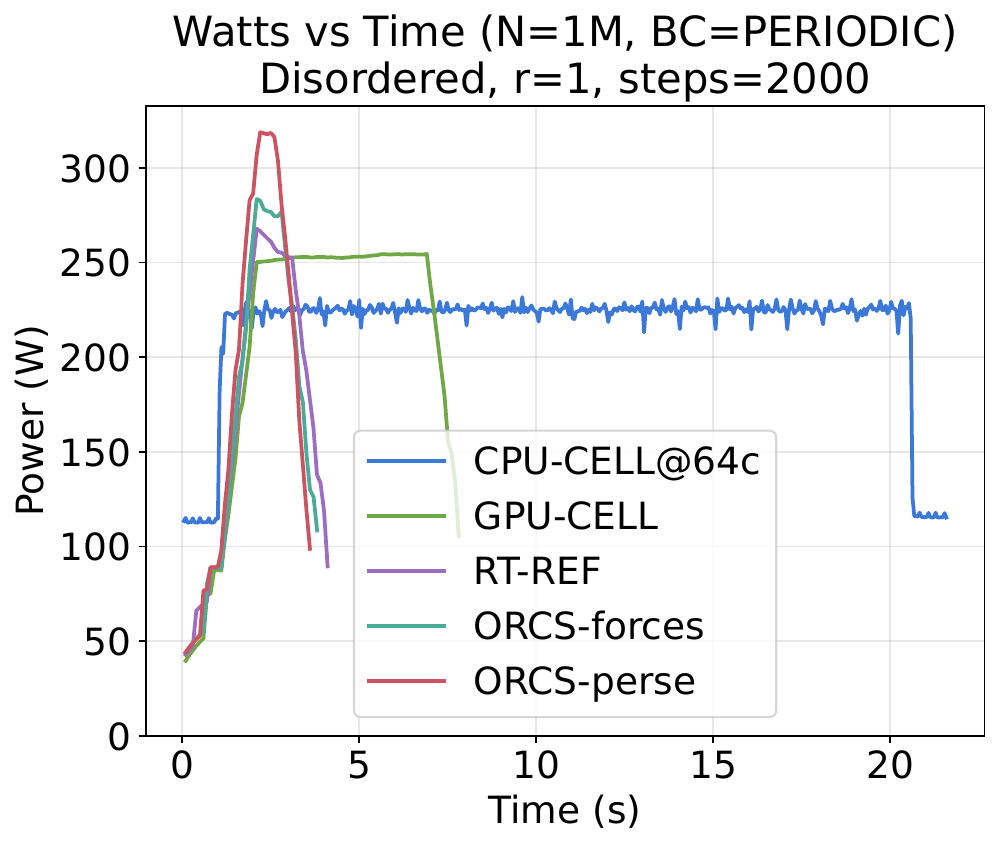}
    \includegraphics[width=0.32\linewidth]{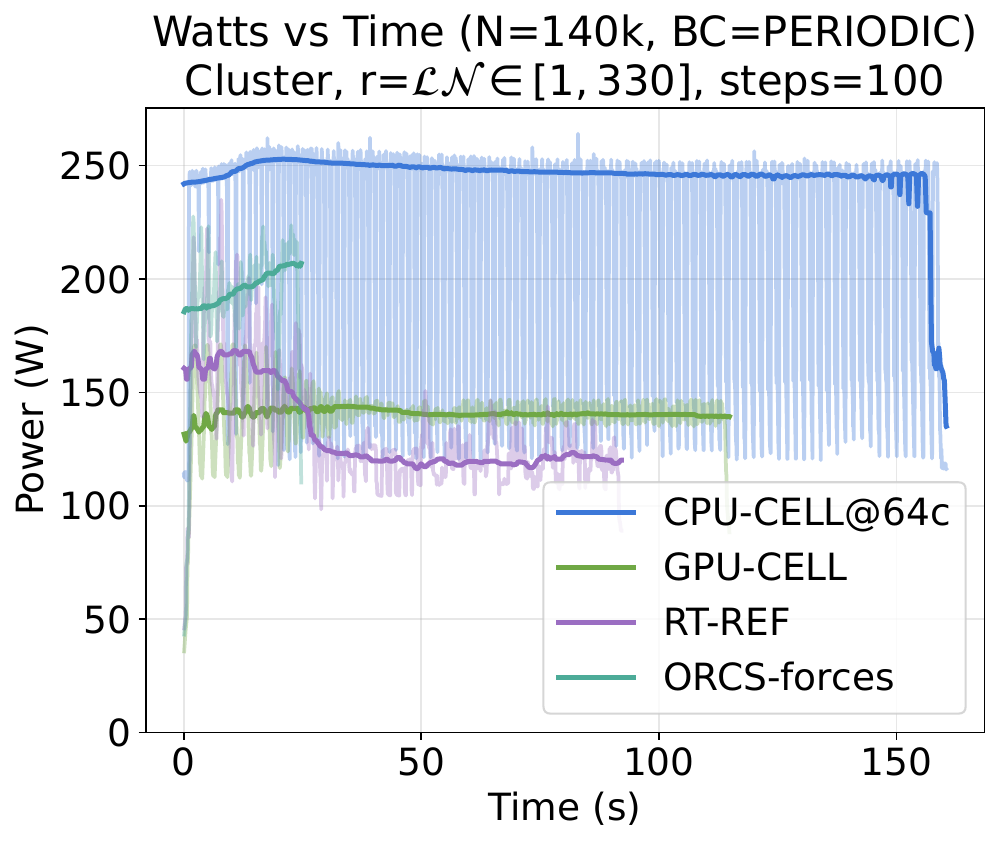}
    \caption{Power consumption time series of all approaches, including \texttt{GPU-CELL} and \texttt{CPU-CELL@64c} for reference.}
    \label{plot:power-ts}
\end{figure*}

For the energy efficiency and scaling benchmarks we have selected three representative cases that exhibit different interaction dynamics, both under wall and periodic BC. These cases are i) Lattice at $r=160$, ii) Disordered at $r=1$ and iii) Cluster at $r=(\mathcal{LN}, r\in [1..330]$).
Figure \ref{plot:power-ts} shows the power consumption time series of these three cases, with the wall and periodic BC in the first and second rows, respectively. The results show that all approaches reach a power consumption that, while not low, remains significantly below the GPU's 600W peak\footnote{We verified that the GPU can reach its peak 600W with CUBLAS GEMM tests.}. This behavior can be explained in great part by the fact that FRNN simulations are often memory-bound. Going column by column we observe that in the Lattice Wall case \texttt{RT-REF} is the highest consuming approach, reaching close to 400W, and followed by \texttt{CPU-CELL@64c} which draws a more stable $\sim 250W$ but sustained for much longer time. The power consumptions of the proposed methods \texttt{ORCS-forces} and \texttt{ORCS-pers\'e} fall in between all the other methods, and \texttt{GPU-CELL} has the lowest power consumption while being competitive. In fact, \texttt{GPU-CELL} is also the fastest under periodic BC. In the second column, Disordered, power consumption curves are much more stable than in Lattice and with not so much differences between wall and periodic BC. Still, one can notice the pulse-like behavior of the RT Core variants, led by \texttt{ORCS-pers\'e} and followed by \texttt{ORCS-forces} and \texttt{RT-REF}. Lastly, the third column, Cluster, presents variance as in Lattice, but the trend lines show stable power consumption except for \texttt{RT-REF} which starts higher and lowers its consumption mid simulation. In this case, it is notorious how \texttt{ORCS-forces} has a much shorter duration roughly at $\sim 200W$ which is less power than \texttt{CPU-CELL@64c}, but higher than \texttt{RT-REF} and \texttt{GPU-CELL}. 

\begin{figure*}[ht!]
    \centering
    \includegraphics[width=0.32\linewidth]{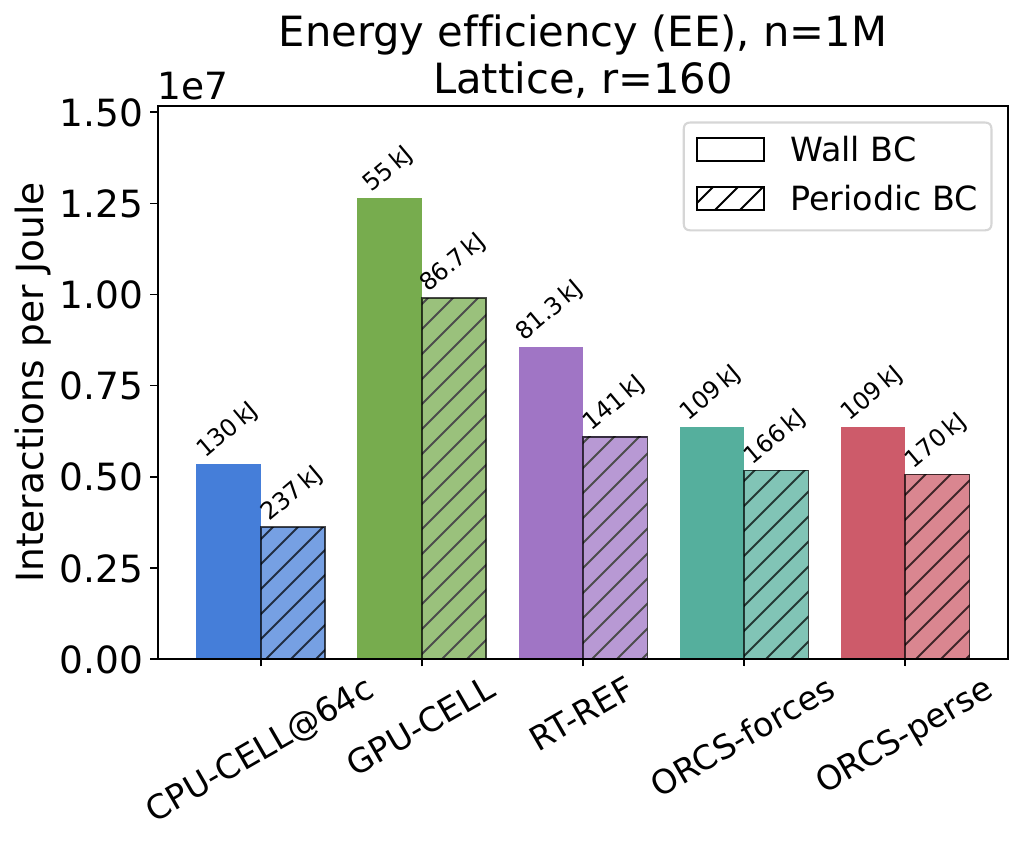}
    \includegraphics[width=0.32\linewidth]{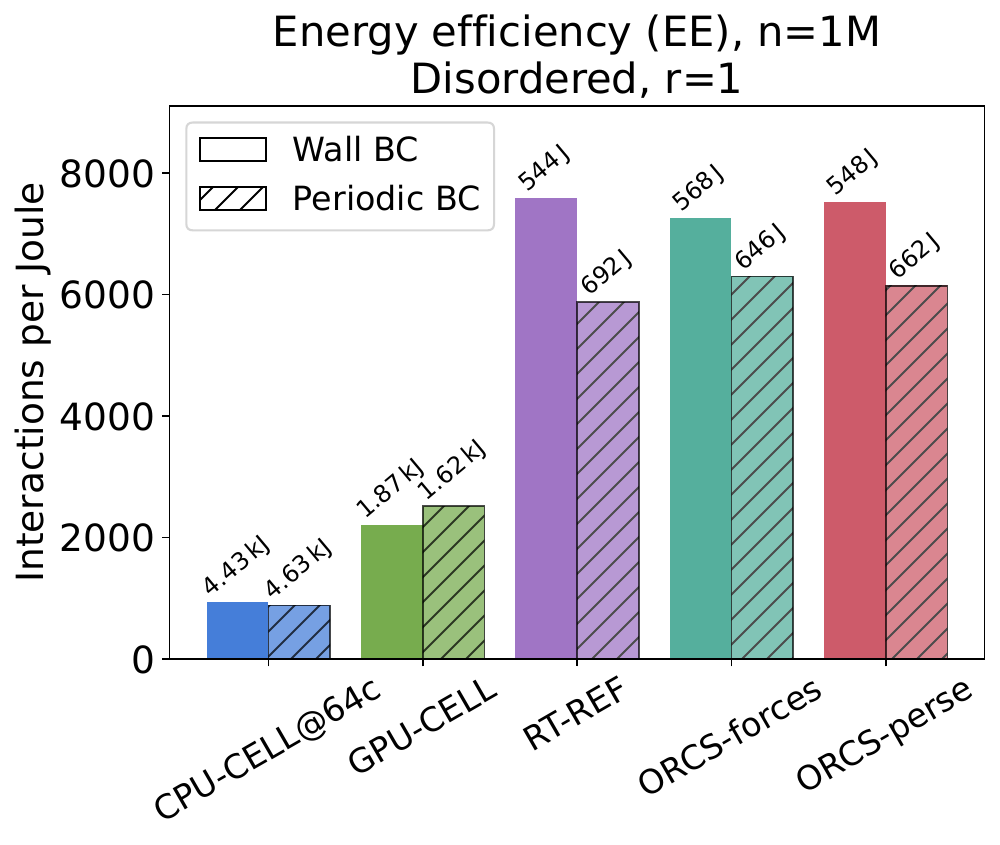}
    \includegraphics[width=0.32\linewidth]{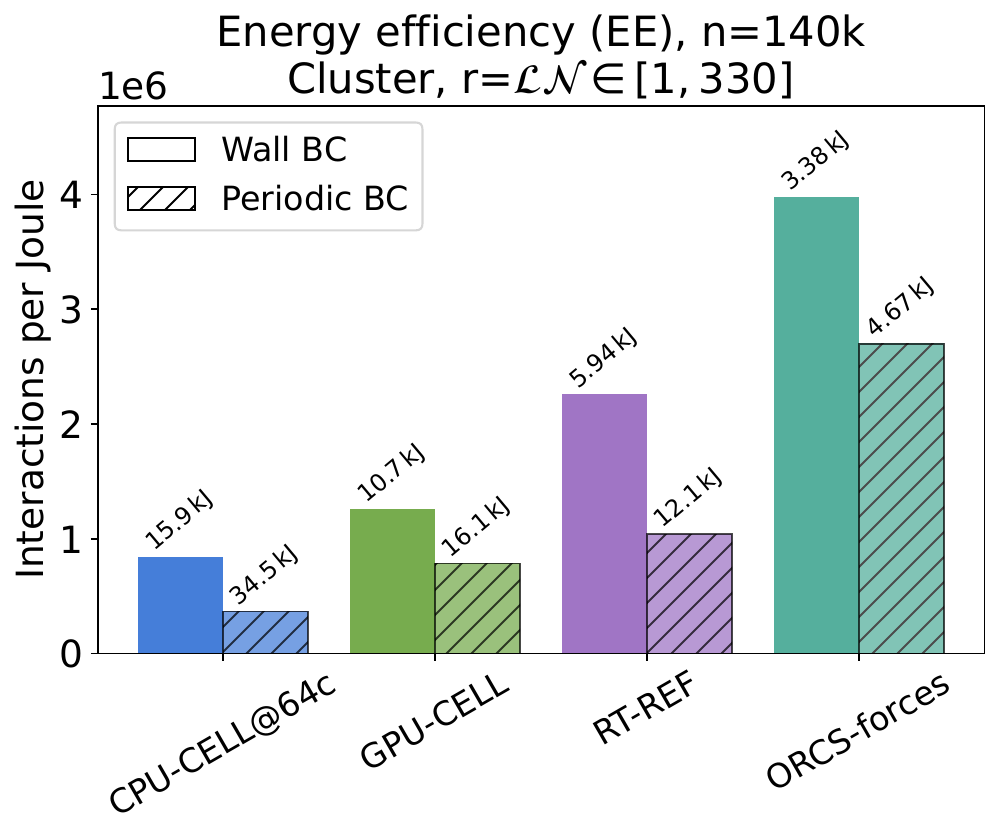}
    \caption{Energy efficiency values for the three selected cases, including \texttt{GPU-CELL} and \texttt{CPU-CELL@64c} for reference.}
    \label{plot:EE}
\end{figure*}

Figure \ref{plot:EE} shows the energy efficiency (EE) of all approaches under the three selected cases. Energy efficiency is defined as interactions per Joule, i.e., 
\begin{equation}
    EE = \frac{I}{E}
\end{equation}
where $I$ is the total number of interactions of the whole simulation, considering $(x_i,x_j)$ and $(x_j,x_i)$ as the same interaction, and $E$ is the total amount of energy (Joules). Additionally, each bar includes the total amount of energy annotated at its top. In the Lattice case, \texttt{GPU-CELL} stands out as the most energy efficient method, reaching up to 12.5M and 1M interactions per Joule for wall and periodic BC, respectively, followed by \texttt{RT-REF} and its proposed variants. The CPU approach, while being the less energy efficient, is still competitive with an EE very close to the proposed variants. In the Disordered case, the RT Core variants all share the highest values of EE, having a significant distance over \texttt{GPU-CELL} and \texttt{CPU-CELL@64c}. RT Core approaches all present similar EE values, although \texttt{RT-REF} is slightly more energy efficient in wall BC, while \texttt{ORCS-forces} is in periodic BC. Lastly, in the Cluster case, we see a progression of EE values from left to right, increasing non linearly and showing that \texttt{ORCS-forces} is the most energy efficient of all, followed by \texttt{RT-REF} but far behind. Again, although the \texttt{CPU-CELL@64c} is the least energy efficient method, it is still competitive and not far from \texttt{GPU-CELL}, making parallel CPU approaches an interesting option both in performance and energy efficiency.

\begin{figure*}[ht!]
    \centering
    \includegraphics[width=0.325\linewidth]{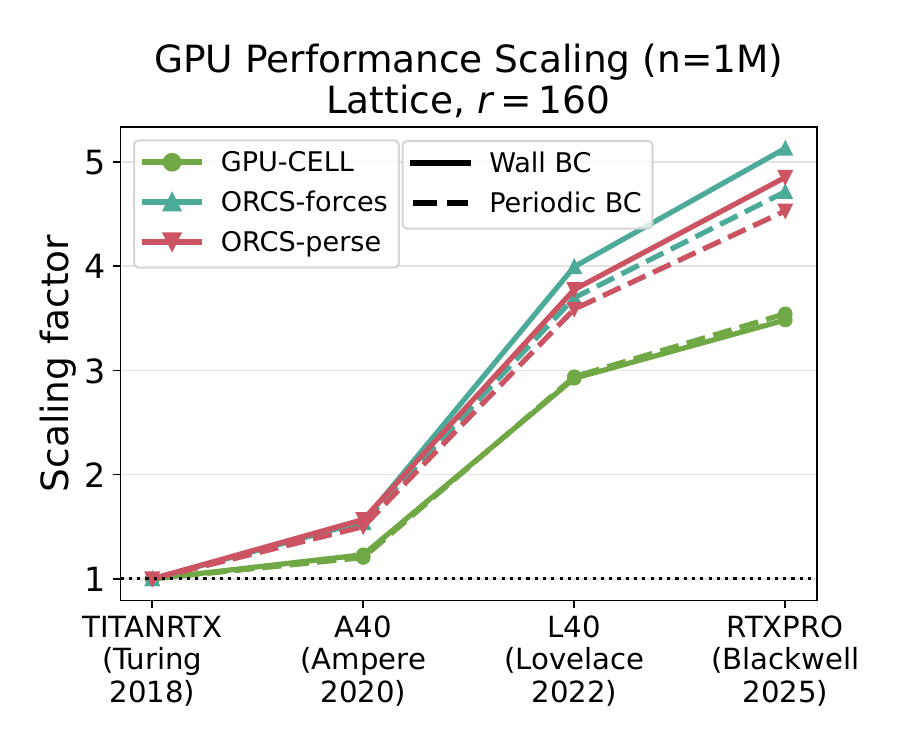}
    \includegraphics[width=0.325\linewidth]{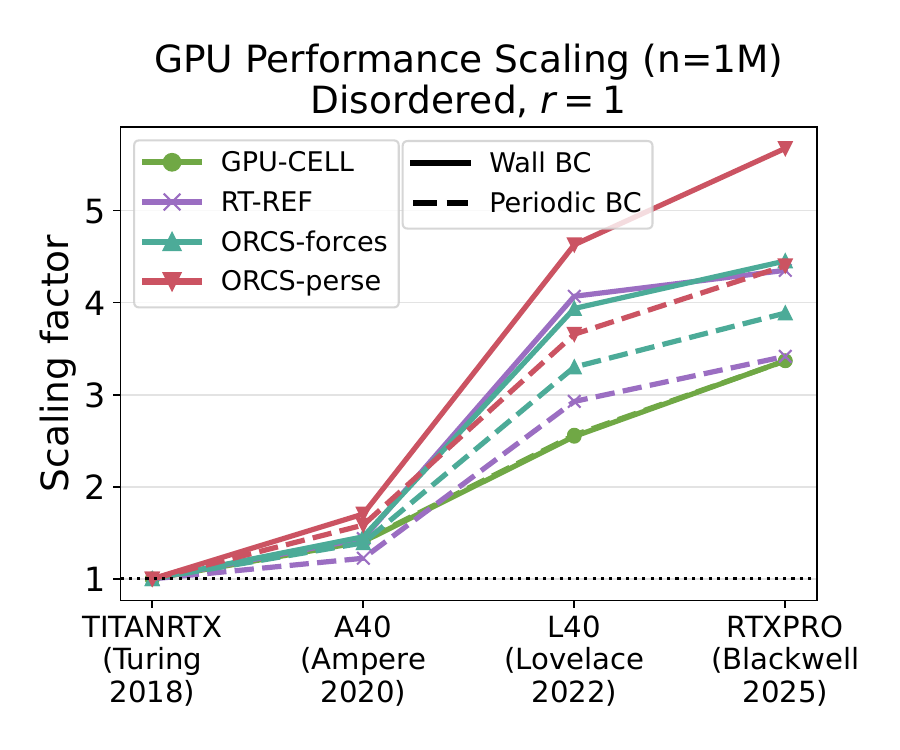}
    \includegraphics[width=0.325\linewidth]{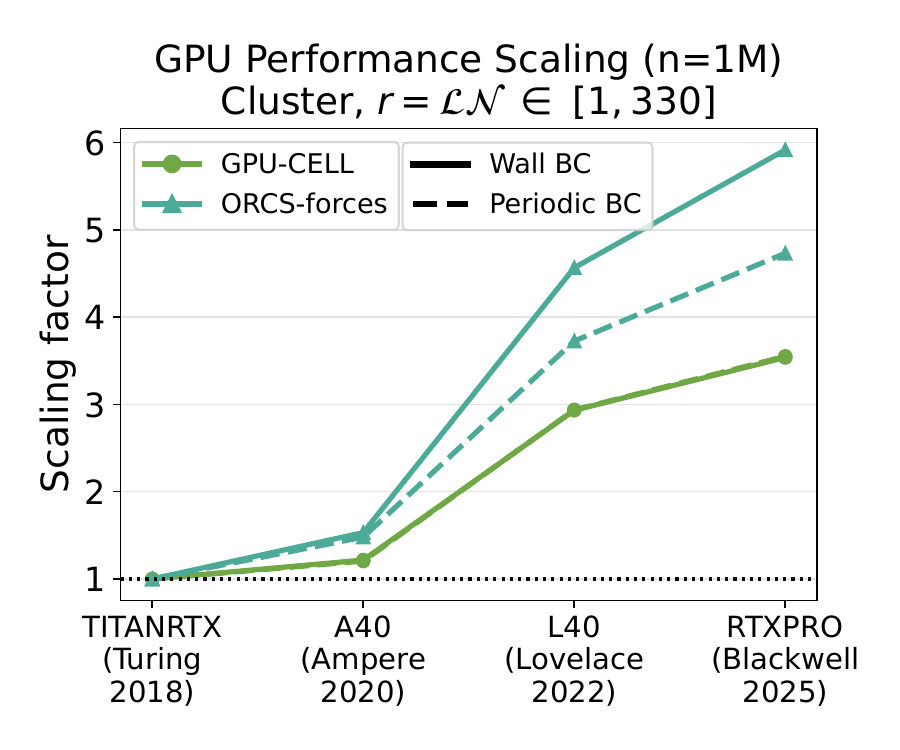}\\
    \includegraphics[width=0.325\linewidth]{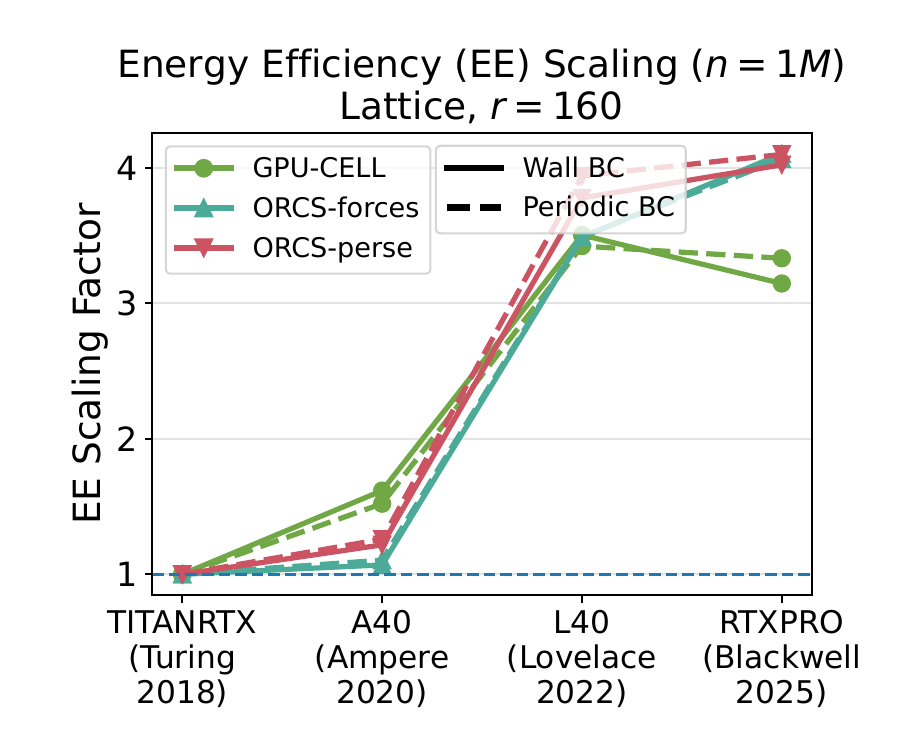}
    \includegraphics[width=0.325\linewidth]{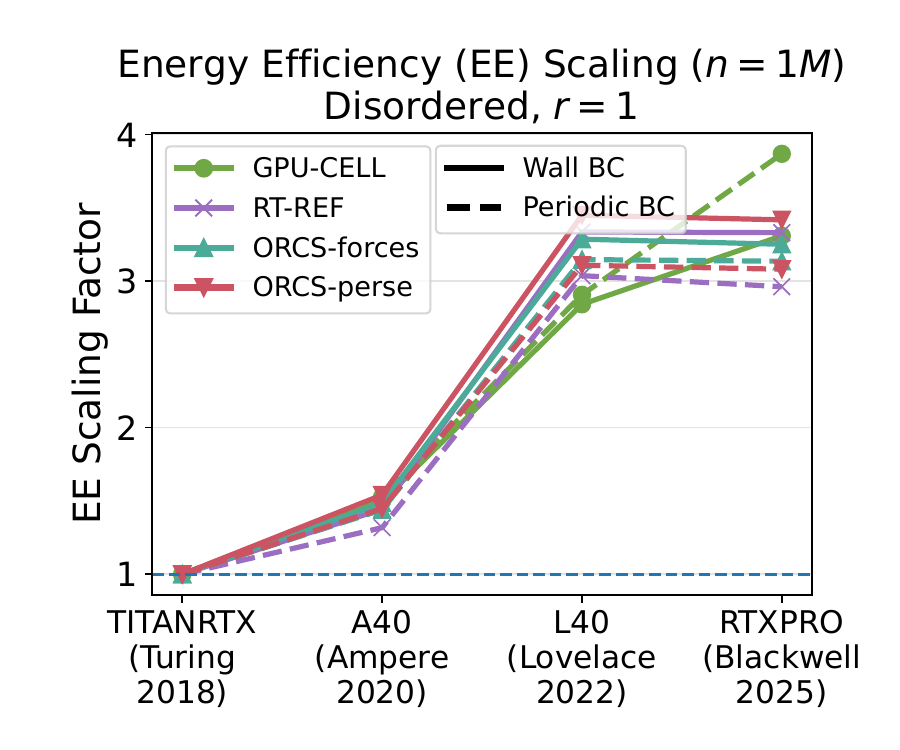}
    \includegraphics[width=0.325\linewidth]{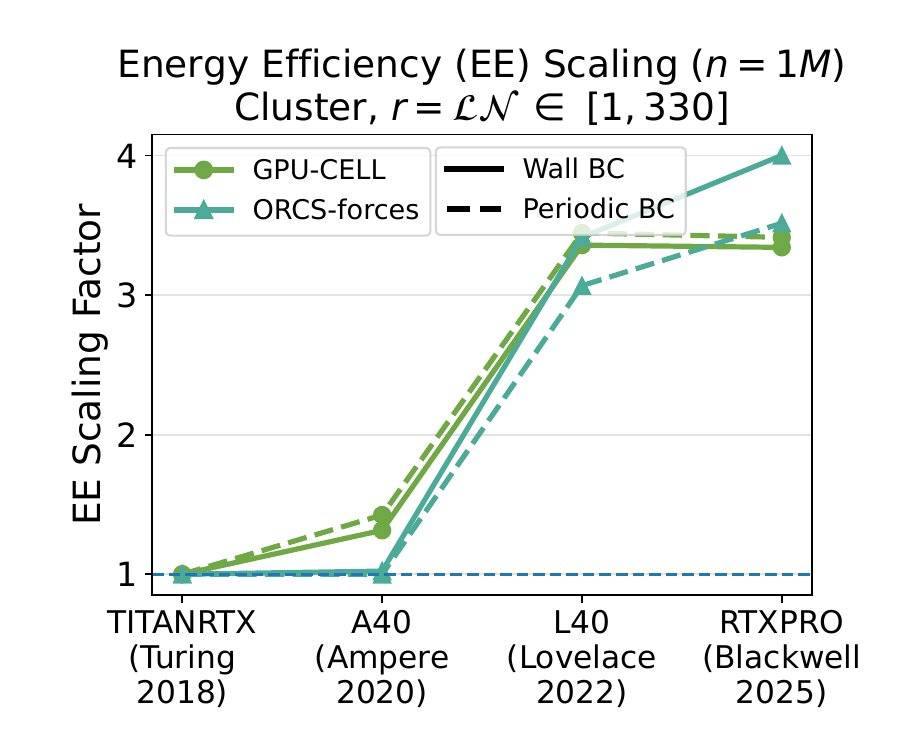}
    \caption{GPU performance scaling (first row) and energy efficiency (EE) scaling (second row) under different particle/radius distributions.}
    \label{plot:scaling-perf}
\end{figure*}

Figure \ref{plot:scaling-perf} presents both the GPU performance scaling and energy efficiency (EE) scaling across the last four GPU generations; from NVIDIA Turing to NVIDIA Blackwell, all with RT Cores. In the first and last column of plots \texttt{RT-REF} approach ran out of memory because it required too many neighbors per particle\footnote{We also tested sizes that would make \texttt{RT-REF} fit even in the TITAN RTX (24GB) for both Lattice ($n=220k$) and Cluster ($n=75k$), but the scaling results on such sizes were misleading, exhibiting under-scaling, as the RTXPRO requires a large problem to saturate its computation.}; Lattice at $r=160$ required 25,000 neighbors per particle and Cluster at $r=\mathcal{LN} \in [1..330]$ requires up to $n$ neighbors per particle (becomes close to a $O(n^2)$ case for the first simulation steps), leaving \texttt{ORCS-forces} and \texttt{ORCS-pers\'e} as the only RT Core candidates. In terms of performance (first row), scaling progressively increases on each GPU generation. In particular, the strongest scaling is found when switching from the A40 (Ampere) to the L40 (Lovelace), followed by the jump from L40 to RTXPRO. Also, the approaches that most scale are the RT Core ones, with the proposed variants being the ones with highest self scaling. In terms of energy efficiency (second row), all approaches scale their EE up to the L40 GPU, with the highest factor found when switching from A40 to L40. On the other hand, the switch from L40 to RTXPRO exhibits mixed results. In Lattice, the \texttt{ORCS} variants did scale their EE when switching from L40 to RTXPRO while \texttt{GPU-CELL} did not, even when its performance did scale up (first row). In the Disordered case the opposite situation is found, all \texttt{ORCS} variants maintain their EE when passing from L40 to RTXPRO, while \texttt{GPU-CELL} further improves it. Lastly, in the Cluster case, \texttt{ORCS-forces} scales its EE significantly, while \texttt{GPU-CELL} maintains it. When performance scales up from one generation to the next while EE does not, it corresponds to cases where the newer GPU chip actually computes faster but at the cost of much more energy usage. In that sense, the Lovelace architecture (L40 GPU) still stands out as highly energy efficient.

\section{Discussion and Conclusions}
\label{sec:conclusions}
Efficient fixed radius nearest neighbors (FRNN) search is a relevant task in fields such as molecular dynamics, materials science, smoothed particle hydrodynamics (SPH), variants of n-body, among many more, as they often require fast simulations of thousands to millions of particles in order to discover and study new phenomena. Using the Lennard-Jones (LJ) potential as a case study, this research has introduced three improvements to the state-of-the-art idea of computing FRNN with ray tracing (RT) cores, presenting relevant results that show when RT core solutions are convenient and when they are not.

The first contribution is \textbf{gradient:} a new adaptive technique that optimizes the ratio of BVH updates/rebuilds in real-time, based on the derivative of a cost function. The \texttt{gradient} approach automatically manages to reduce the rate of updates per rebuild when the simulation has stronger dynamics, and automatically increases this rate when the dynamics are slower. As a whole, the gradient scheme makes the RT core up to $\sim3.4\times$ faster than other known approaches such as rebuilding at a fixed rate, which cannot adapt in real-time, or rebuilding once the average cost of simulating with updates surpasses the cost of just rebuilding. Regardless of the specific speedup values vs the fixed approach selected, the greatest strength of \texttt{gradient} is its ability to adapt by finding the optimal update/rebuild ratio in real-time on its own, without requiring the programmer to exhaustively test a list of different fixed ratios. 

The second contribution is \textbf{ORCS} which is the idea of computing FRNN with RT cores without neighbor lists. Two variants where proposed and evaluated; the first one is \texttt{ORCS-pers\'e}, which can run the whole simulation within the ray tracing pipeline for simulations where all particles have the same radius. The second one is \texttt{ORCS-forces} which accumulates the force contribution immediately into the particle's final force for each neighbor interaction triggered by the ray tracing pipeline, and then uses a CUDA kernel to displace the particles accordingly. This variant supports particles with variable radius. Experimental results show that the speedup obtained by the ORCS variants over a highly parallel CPU approach (\texttt{CPU-CELL@64c}) can be higher than what a base RT method (\texttt{RT-REF}) obtains, in particular when the radius is small or follows log normal distributions. %There were however where the proposed variants were not faster, such as when a significant number of particles have large radius. 
%For instance, at low radius with Lattice distributions \texttt{ORCS-pers\'e} can increase the base RT core speedup from $18\times$ to $24\times$. In the Cluster case \texttt{ORCS-forces} not only manages to reach up to $\sim 40\times$ of speedup over the reference CPU approach, but it also the only RT core approach capable of handling such varied  neighbor interactions without running out of memory, as it does not require a neighbor list. 
Not all cases were favorable though; when a significant number of particles have large radius, the base RT core method runs faster than the proposed variants. Furthermore, for clustered particle distributions with common large radii, RT core approaches become inefficient as the number of interactions is too large for ray tracing to provide a benefit. For such cases, a parallel CPU or GPU cell list approach still seems more convenient.

The third contribution is \textbf{Ray Traced Periodic Boundary Conditions}; by launching additional rays with offsets for the particles near the boundaries, the simulation manages to capture the interactions from particles from the opposite walls, without introducing any additional compute kernel. Experimental results show that this idea proposed does not introduce any significant penalty to the performance of the RT core methods, in fact it reaches higher speedups compared to the Cell based approaches both on CPU and GPU. The core idea for supporting Periodic BC is a research problem by itself that could be further improved, as it is currently based on the largest radius of the system. This works well for particles with equal or similar radius, but it may introduce unwanted performance penalty if one particle has a radius much larger than the others.

Energy efficiency (EE) results were also computed, in the form of interactions per Joule. The results have shown that the RT core solutions are the most energy efficient ones at small radius, and for log normal radius the proposed \texttt{ORCS-forces} variant is significantly more energy efficient than the base known RT core idea and all other approaches. On the contrary, when many particles have large search radius, the classic \texttt{GPU-CELL} approach is the most energy efficient one, as the number of interactions is too large for RT cores to discard work.

An experimental evaluation of performance and EE across different GPU generations was also presented; from NVIDIA 2022 Turing (TITANRTX) to 2025 NVIDIA Blackwell (RTX Pro 6000 Blackwell Server Edition). Scaling results have shown that the scaling factor from Ampere (A40) to Lovelace (L40) is the strongest one, gaining up to double of the performance and energy efficiency for some approaches. In some cases, several approaches just maintain the EE and in one particular case, the Lattice at $r=160$, \texttt{GPU-CELL} scales down its EE from L40. These scaling results show a relevant trend change starting from Blackwell, which is that the performance gains of new high end GPUs are not necessarily bringing an improvement in EE. In fact, the peak 600W power consumption of the Blackwell GPU supports this conclusion, compared to the 300W of the L40 GPU. We expect that future GPU generations return to the EE scaling trend that the Lovelace architecture provided for FRNN computations.

Future work can extend \texttt{gradient} to optimize towards energy efficiency, i.e., to optimize the update/rebuild scheme according to the energy and power consumption values measured in real time through a tool like NVIDIA NVML, or equivalent, instead of using performance timers. Another line of work is to leverage new architectural RT features for FRNN. For example, shader execution reordering (SER) might bring some additional performance gains by sorting rays during execution; research would focus on exploring if the benefit surpasses the sacrifice in performance because of the sorting itself. Lastly, the current method for supporting periodic BC has a potential performance penalty when one or few particles have a very large radius, making the RT cores launch unnecessary rays. We hope that a deeper research on handling periodic BC with RT cores will lead to more efficient ideas.

\section*{Acknowledgements}
This research was supported by the ANID FONDECYT grants \#1221357 and \#1241596, ANID ECOS \#230017 and the Patag\'on Supercomputer of Austral University of Chile (FONDEQUIP \#EQM180042). Additional funding was also provided by the French National Research Agency (ANR-20-CE46-0004), the ECOS-ANID project C23E03, the Region Nouvelle-Aquitaine and the Ceramics \& ICT Graduate School (U. of Limoges).

\bibliographystyle{elsarticle-num}
\bibliography{main}
\end{document}